\documentclass[12pt]{article}
\usepackage{url}

\usepackage[utf8]{inputenc}
\usepackage{cite}
\usepackage{amsfonts, amsbsy, amssymb} 
\usepackage{amsthm}
\usepackage{amsmath}
\usepackage{color}
\usepackage{float}
\usepackage{subfigure}
\usepackage{graphicx}
\usepackage{bm}
\usepackage[normalem]{ulem}
\usepackage{authblk}
 \usepackage{geometry}
\begin{document}

\title{Revealing Roaming on the Double Morse Potential Energy Surface with Lagrangian Descriptors}
\author[a]{Francisco Gonzalez Montoya}
\author[a]{Stephen Wiggins}
\affil[a]{ School of Mathematics, University of Bristol, Bristol, BS8 1UG, United Kingdom }
\maketitle

\begin{abstract}
In this paper, we analyse the phase space structure of the roaming dynamics in a 2 degree of freedom potential energy surface consisting of two identical planar Morse potentials separated by a distance. This potential energy surface was previously studied in \cite{carpenter2018dynamics}, and it has two potential wells surrounded by an unbounded flat region containing no critical points. We study the phase space mechanism for the transference between the wells using the method of Lagrangian descriptors.

\end{abstract}

\begin{center}
francisco.glz.mty@gmail.com

\end{center}
\begin{center}
s.wiggins@bristol.ac.uk
\end{center}

\section{Introduction}
\label{sec:intro}

The purpose of this paper is to reveal the phase space mechanism for roaming in the double Morse potential energy surface using the method of Lagrangian descriptors. The roaming mechanism for chemical reactions was first studied in this model in \cite{carpenter2018dynamics}. 

The roaming mechanism for chemical reactions was reported in \cite{townsend2004roaming}. This paper inspired and motivated further work on the topic that is continuing. Some recent reviews of the topic are  \cite{Bowman2011Suits,Bowman2011roaming,BowmanRoaming,suits2008, Mauguiere2017}.

Following \cite{carpenter2018dynamics}, a concise description of the roaming mechanism that seems to capture all such reactive phenomena attributed to roaming to date is the following. An energised molecule, which appears on the verge of dissociating into two fragments, instead exhibits behaviour where one fragment begins to rotate about the other in the flat region of the potential energy surface, culminating in the two fragments re-encountering each other in a ``reactive event''. This reactive event may be constitutionally identical to one for which a familiar transition state associated with an index one saddle can be identified, but the roaming reaction path is observed to avoid the intrinsic reaction path.

The double Morse potential energy surface allows us to analyse key feature of the reaction dynamics which appear to underlay the roaming mechanism. In particular, the double Morse has a flat ``roaming region'' where it can be rigorously proven that there are no saddle points. For explaining reaction mechanisms in the context of the potential energy surface index one saddles have been a fundamental building block. This has motivated efforts to locate index one saddle points that can be attributed as the mechanism for roaming (``roaming saddles'').

However, the work in \cite{carpenter2018dynamics} showed that roaming mechanism in the double Morse potential occurs without the influence of saddle points. In particular, it was demonstrated that the roaming was governed by phase space structures governing dynamics in the flat region of the potential energy surface. These phase space structures are normally hyperbolic invariant manifolds (NHIMs) \cite{Wiggins_book1994,Wiggins2001,Uzer2002, Wiggins08,Wiggins2016}, which in the case of 2 degree of freedom Hamiltonian systems are unstable periodic orbits \cite{Mauguiere2015}.

The main issue of interest in \cite{carpenter2018dynamics} was the motion of trajectories between the two wells and the intermediary dynamics in the flat region. It was conjectured that three types of unstable periodic orbits mediated the dynamics. The description of roaming given in \cite{carpenter2018dynamics} was based on the behaviour of trajectories initiated on periodic orbit dividing surfaces associated with the three types of periodic orbits. The trajectory behaviour confirmed the conjecture. However, this work did not examine the phase space structures that determine the different types of trajectory behaviour. Such an analysis would involve an understanding of the geometry of the stable and unstable manifolds of the periodic orbits in phase space, similar in spirit to the analysis that has been carried out in \cite{krajvnak2018phase, krajvnak2018influence}. 

In this paper, we will determine the phase space pathways for the roaming mechanism in the double Morse oscillator using the method of Lagrangian descriptors. Lagrangian descriptors are a trajectory diagnostic that has proven useful for revealing phase space structures in dynamical systems.  The method was originally developed to reveal fluid flow structures in Lagrangian transport problems (hence the name Lagrangian descriptors) \cite{chaos}.  However, in recent years, the ease of implementation and interpretation of the method has applied in fields far beyond fluid mechanics. In particular, in theoretical chemistry the method has seen a diverse collection of applications to problems in chemical reaction dynamics in recent years, see, e.g., \cite{patra2018detecting, craven2016deconstructing, craven2017lagrangian, craven2015lagrangian, junginger2016transition, revuelta2017transition, junginger2017chemical, feldmaier2017obtaining, JH2015}.

The method facilitates a high-resolution approach for exploring high dimensional phase space with low dimensional sets.  It can also be applied to both Hamiltonian and non-Hamiltonian systems \cite{lopesino2017} as well as to systems with arbitrary, even stochastic, time-dependence \cite{balibrea2016lagrangian}. Moreover, Lagrangian descriptors can be applied directly to trajectory data sets, without the need for an explicit dynamical system \cite{mmw14}.

The ``classical" Lagrangian descriptor field is computed by selecting a region of phase space and then choosing a grid of initial conditions for trajectories in this set. Points in the set are assigned a value according to the arc length of the trajectory starting at that initial condition after integration for a fixed, finite time, both backward and forward in time (all initial conditions in the grid are integrated for the same time). The idea underlying Lagrangian descriptors is that the influence of phase space structures on trajectories will result in differences in arc length of nearby trajectories. This has been verified and quantified in terms of the notion of singular structures in the Lagrangian descriptor fields, which are easy to recognise visually \cite{Mancho2013, lopesino2017}.

Trajectories are the elementary objects that are used to explore phase space.  In fact, phase space structure is built from trajectories and their behaviour. For high dimensional phase space this approach is problematic and prone to issues of interpretation since a tightly grouped set of initial conditions may result in trajectories that become far with respect to each other in high dimensional phase space. The method of Lagrangian descriptors takes a new and different approach by encoding the phase space structure in the initial conditions of trajectories, rather than the precise location of their futures and pasts (after a specified amount of time). Hence, a low dimensional set of phase space can be selected and sampled with a grid of initial conditions of high resolution. Since the phase space structure is encoded in the initial conditions of the trajectories, no resolution in the chosen slice of phase space is lost as the trajectories evolve in time.

This paper is outlined as follows. In section \ref{sec:DMPES}, we recall the main features of the double Morse potential energy surface and resulting Hamiltonian system as described in \cite{carpenter2018dynamics} that are relevant to our analysis.  In section \ref{sec:config}, we define the features of the double Morse potential energy surface that are relevant to our analysis.  In section \ref{sec:LD} we describe Lagrangian descriptors. In section \ref{sec:PO}, we show how Lagrangian descriptors can be used to compute periodic orbits and their stable and unstable manifolds. In section \ref{sec:roaming}, we describe the phase space pathways for roaming using the stable and unstable manifolds of the periodic orbits. 

\section{The double Morse potential energy surface}
\label{sec:DMPES}

The 2 degree of freedom Hamiltonian system that we consider is the sum of two identical planar Morse potentials \cite{Morse29} separated by a distance $2b$:

\begin{eqnarray}
V (x, y) & = & D_e \left(1 - \exp\left(-\sqrt{\frac{k}{D_e}}  (\sqrt{(x-b)^2 + y^2} - r_e)\right)  \right)^2 \nonumber \\
&  + & 
D_e \left(1 - \exp\left(-\sqrt{\frac{k}{D_e}}  (\sqrt{(x+b)^2 + y^2} - r_e)\right)  \right)^2 -D_e.
\label{eq:2Morse_pot}
\end{eqnarray}

\noindent
and kinetic energy of the form:

\begin{equation}
T(p_x, p_y) = \frac{p_x^2}{2m} + \frac{p_y^2}{2m}.
\label{eq:KE}
\end{equation} 

\noindent
Therefore the Hamiltonian is:

\begin{equation}
H(x, y, p_x, p_y) = \frac{p_x^2}{2m} + \frac{p_y^2}{2m} + V (x, y).
\label{eq:Ham}
\end{equation}

\noindent
Following \cite{carpenter2018dynamics}, the parameters for the Hamiltonian are chosen as follows:

\begin{equation}
D_e = 100, \, \, r_e =1, \, \, k=200, m=1, b=3.
\label{eq:params}
\end{equation}

\noindent The parameter $D_e$ is the depth of the Morse potential well. Therefore for total energies above $D_e$ trajectories can dissociate, i.e. they become unbounded, and the value of this parameter is referred to as the dissociation threshold. Taking the units of energy to be $\rm kcal \, mol^{-1}$, and units of distance to be \AA, the parameters of the single Morse potential functions in equation \eqref{eq:2Morse_pot} correspond, roughly, to the vibration of a CH bond. Chemically, this potential could be considered to represent an atom interacting with a diatomic molecule, or, mechanically, a rigid dumbbell interacting with a particle.  The equilibrium bond length between the two atoms of the diatomic is set by $b$, and the bond length of the third atom to one of the atoms of the diatomic is set by $r_e$.

\section{Features of the double Morse potential energy surface}
\label{sec:config}

%figure1
\begin{figure}
\begin{center}
\scalebox{0.7}{\includegraphics{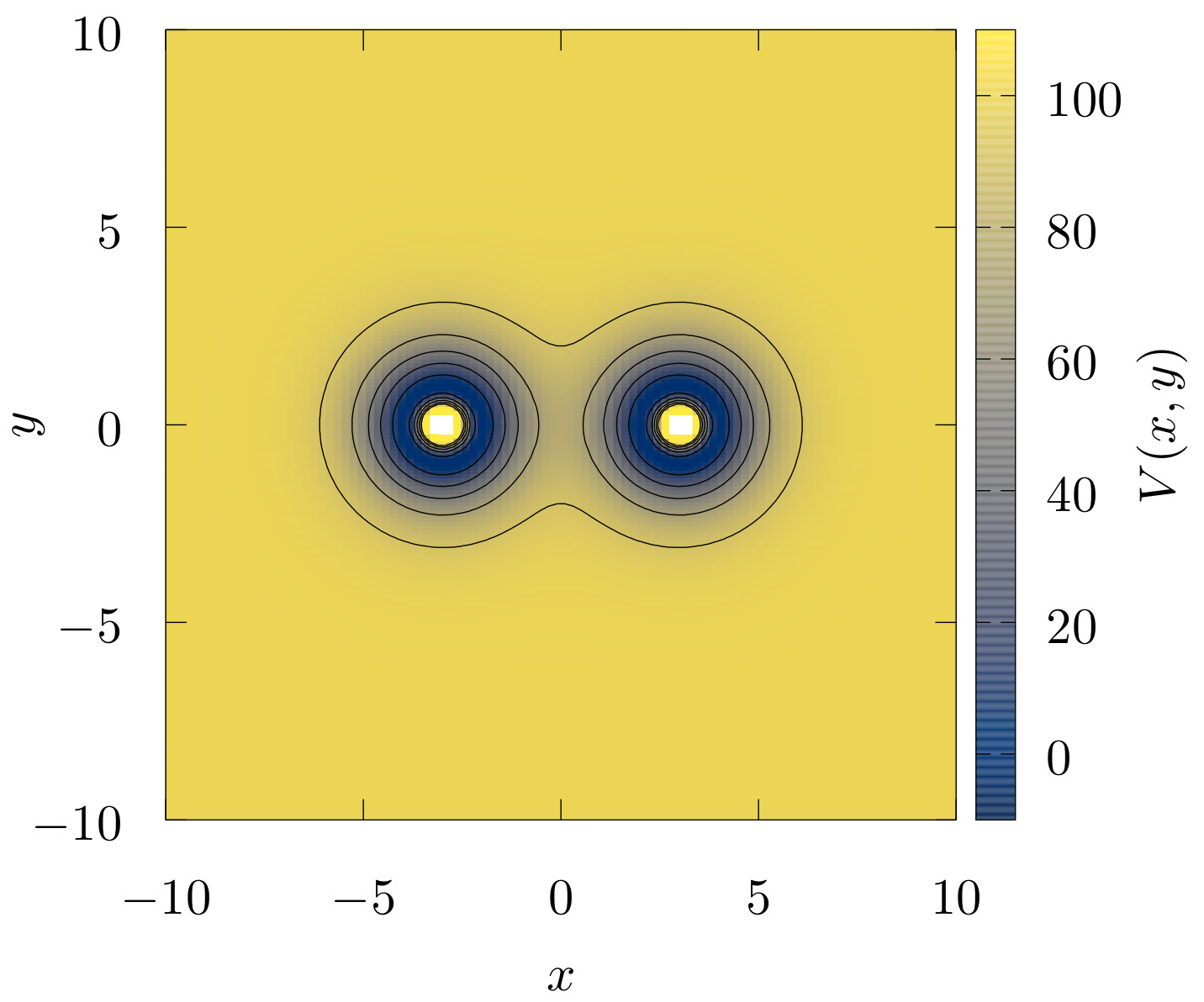}}
\caption{Double Morse potential energy in colour scale. The contour lines black are the equipotential lines for different values of the potential energy. The potential has two wells and two forbidden regions, each one in the centre of each well. The potential energy converges to a constant value in the asymptotic region. 
\label{fig:V_c}}
\end{center}
\end{figure}

%figure2
\begin{figure}
\begin{center}
\scalebox{0.7}{\includegraphics{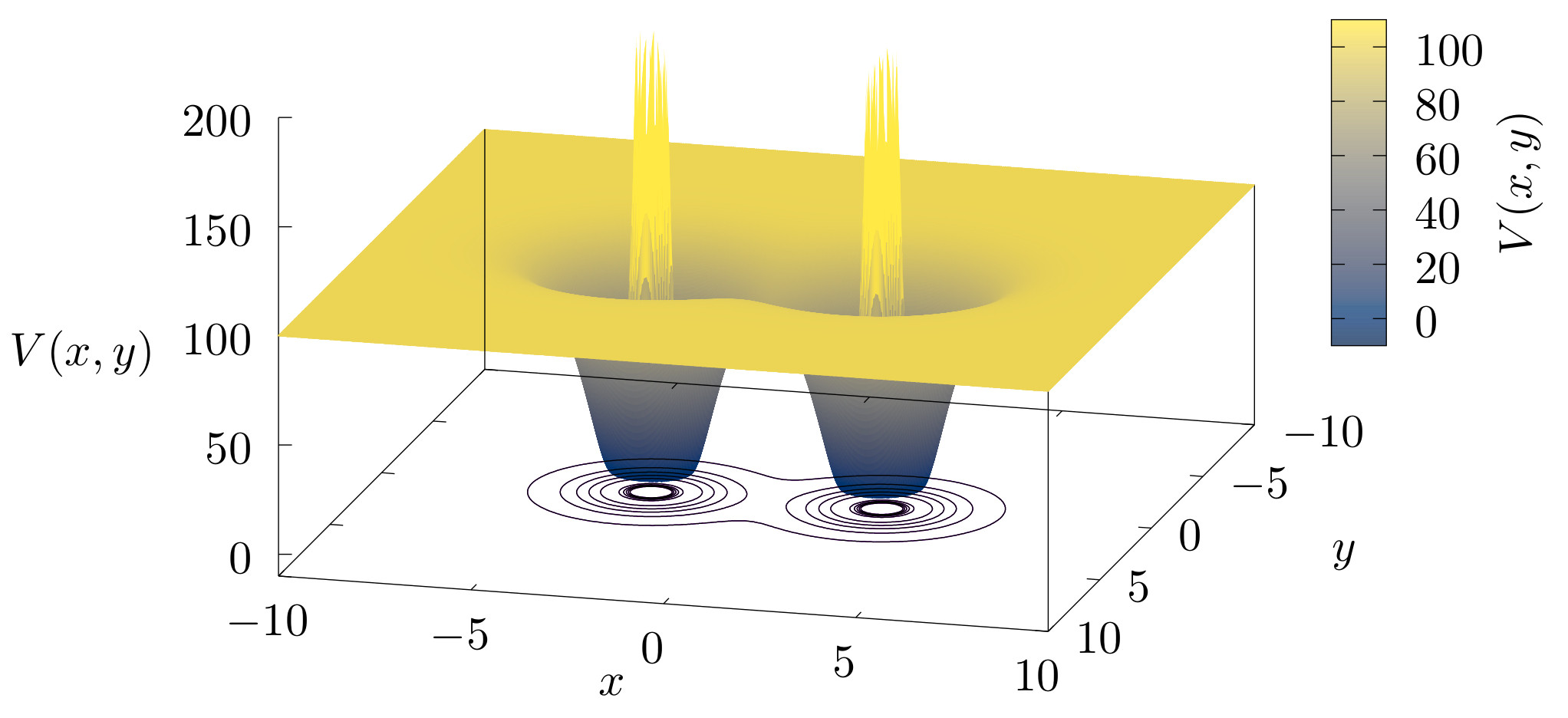}}
\caption{Double Morse potential energy surface in colour scale. The potential has a saddle point between the two wells.
\label{fig:V}}
\end{center}
\end{figure}

%figure3
\begin{figure}
\begin{center}
\scalebox{0.6}{\includegraphics{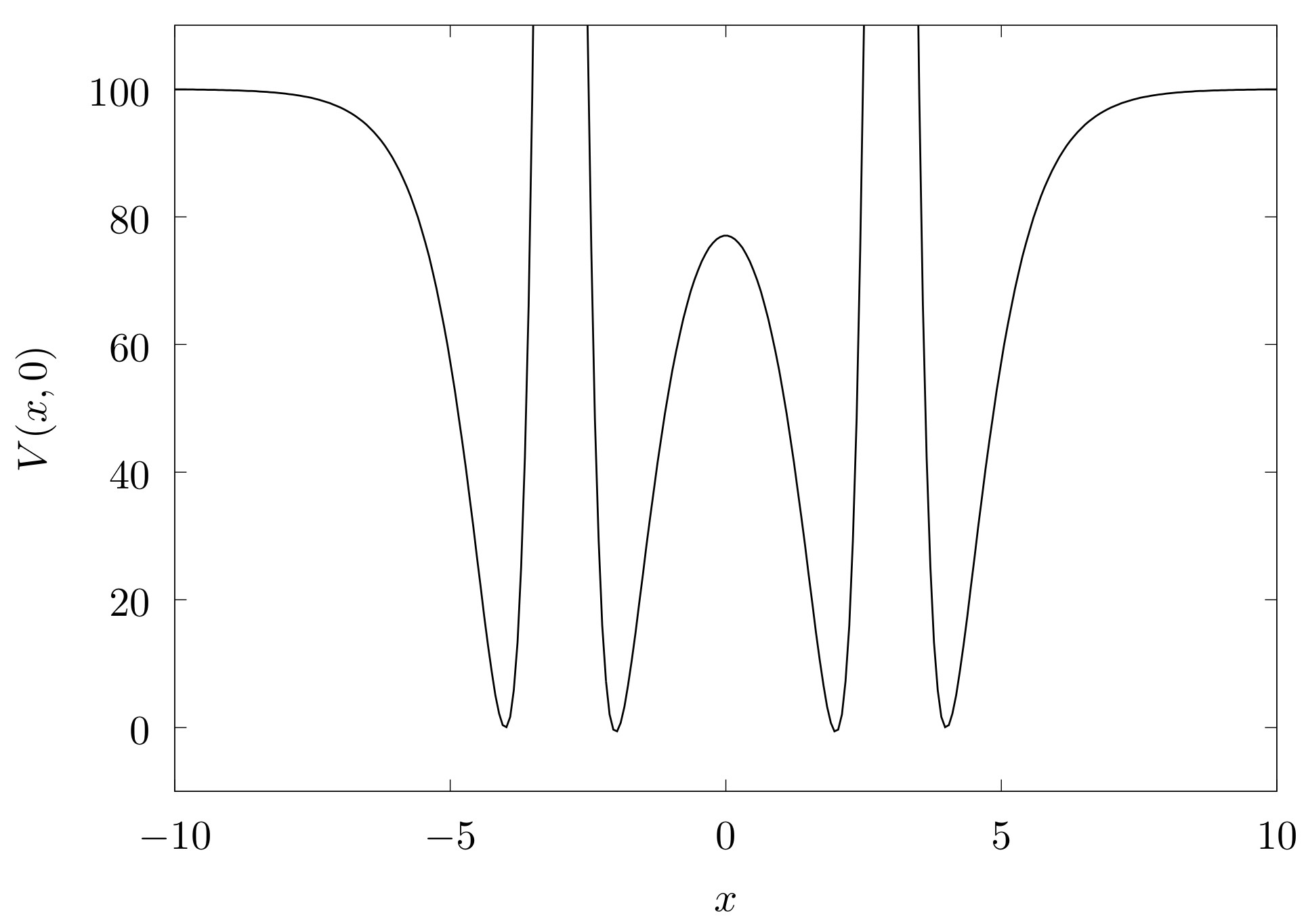}}
\caption{Double Morse potential energy evaluated on the line $y=0$. The energy of the saddle point at the origin $E_s$ is below 80.
\label{fig:V(x,0)}}
\end{center}
\end{figure}

The figures \ref{fig:V_c}, and \ref{fig:V} show plots of the potential energy in colour scale and equipotential lines. At the origin, $(x=0, y=0)$ the potential energy has an index one saddle. The potential energy evaluated in the saddle point is $V(0,0) = E_{s}$. This index one saddle point exists for all values of $b$, and $E_{s}$ is below $D_e$ in this example, see Fig. \ref{fig:V(x,0)}.

The double Morse potential energy function goes exponentially fast to zero in the asymptotic flat region. This almost flat region plays an essential role in the roaming phenomenon, and it is called the {\em roaming region}. 

\section{Lagrangian descriptors and 2 degree of freedom Hamiltonians systems} 
\label{sec:LD}

The ``classical'' Lagrangian descriptor is based on the arc length of a trajectory. In the present article, we are going to use a variant based on the integral of the square of the infinitesimal arc length proposed in \cite{lopesino2017}. Moreover, the present variant considers two modification in the definition of the Lagrangian descriptor. The purpose of the first modification is to differentiate between the phase space structures defined with the propagation of the trajectories forward and backwards in time. The second modification in the definition of Lagrangian descriptor allows the different time propagation of the trajectories for different initial conditions in order to study the phase space structures contained in a specific region of the phase space. We consider these modifications in order to improve the visibility and reveal the important phase space structures in the present study of roaming and dissociation. The definition of this variant of the Lagrangian descriptor is as follows. 

Let us consider the autonomous ordinary differential equations

\begin{equation}
\frac{d \mathbf{x}}{dt} = \mathbf{v}(\mathbf{x}), \quad \mathbf{x} \in \mathbb{R}^n \;,\; t \in \mathbb{R}
\end{equation}

\noindent where $\mathbf{v}(\mathbf{x},t) \in C^r$ ($r \geq 1$) in $\mathbf{x}$ and continuous in time. The definition of Lagrangian descriptor depends on the initial condition $\mathbf{x}_{0} = \mathbf{x}(t_0)$, on the time interval $[t_0+\tau_{-},t_0+\tau_{+}]$, and takes the form,

\begin{eqnarray}
M(\mathbf{x}_{0},t_{0},\tau_{+},\tau_{-}) & = & M_{+}(\mathbf{x}_{0},t_{0},\tau_{+}) - M_{-}(\mathbf{x}_{0},t_{0},\tau_{-}) \nonumber  \\
& = & 
\displaystyle{ \int^{t_{0}+\tau_{+}}_{t_{0}} \sum_{i=1}^{n} |\dot{x}_{i}(t;\mathbf{x}_{0})|^2 \; dt - \int^{t_{0}}_{t_{0}+\tau_{-}} \sum_{i=1}^{n} |\dot{x}_{i}(t;\mathbf{x}_{0})|^2 \; dt ,} \label{eq:LD}
\end{eqnarray}

\noindent where the over dot symbol represents the derivative with respect to the time $t$ and $\tau_{+} \geqslant 0$ and $ \tau_{-} \leqslant 0 $ are freely chosen parameters. In the ``classical'' Lagrangian descriptor the magnitude of $\tau_{+}$ and $\tau_{-}$ are equal for all points $x_0$. In the present variant of Lagrangian descriptor, the values of $\tau_{+}$ and $\tau_{-}$ can change between different initial conditions. This modification allows us to stop the integration once a trajectory leaves a specific region in the phase space and reveal only the structures related contained in this region.

The phase space of a 2 degree of freedom Hamiltonian system has 4 dimensions. Considering the conservation of the energy, it is possible to represent the dynamics of the system in the 3-dimensional constant energy manifold. In the constant energy manifold, it is possible to visualise the dynamics and identify the essential structures to understand the dynamics.

The periodic orbits are essential structures to understand the dynamics of the system in the constant energy manifold. Around a stable periodic orbit, the KAM-tori confine the trajectories in the region defined by the tori. In contrast, the dynamics in a neighbourhood of an unstable hyperbolic periodic orbit has a different nature. Its stable and unstable manifolds intersect in the hyperbolic periodic orbit and direct the trajectories in the neighbourhood. The definition of the stable and unstable manifolds $W^{s/u} (\Gamma)$ of the periodic unstable orbit $\Gamma$ is the following,

\begin{equation}
 W^{s/u} (\Gamma) = \lbrace  \mathbf{x}  \vert \mathbf{x}(t)  \rightarrow  \Gamma, t \rightarrow \pm \infty \rbrace.
\end{equation}

In other words, the stable manifold $W^s(\Gamma)$ is the set of trajectories that converge to the periodic orbit $\Gamma$ as the time $t$ goes to $\infty$. The definition for the unstable manifold $W^u(\Gamma)$ is analogous, that is, it is the set of trajectories converging to the periodic orbit as the time $t$ goes to $-\infty$. 

A 2 degree of freedom Hamiltonian system, $W^{s/u} (\Gamma)$ has 2 dimensions and form impenetrable barriers in the constant energy manifold. The segments of the stable and unstable manifolds direct the transport between different regions \cite{Kovacks2001}.  

Another important property of the stable and unstable manifolds related to the chaotic dynamics is that, if a stable manifold and an unstable manifold intersect transversally at one point, then an infinite number of transversal intersections between them exist. The structure generated by the union of the stable and unstable manifolds is called tangle and defines a set of tubes that direct the dynamics in the phase space. The trajectories in a tube never cross the boundaries of a tube. This fact is a consequence of the uniqueness of the solution of the ordinary differential equations. 

The Lagrangian descriptors are convenient tools to explore the phase space structure, in particular, to find stable and unstable manifolds of periodic orbits. To understand the basic idea that underpins the detection, let us consider the behaviour of the trajectories in a neighbourhood of stable manifold $W^{s}(\Gamma)$. The trajectories in $W^{s} (\Gamma)$ converge to the periodic orbit $\Gamma$, and the nearby trajectories in the neighbourhood of $W^{s} (\Gamma)$ have similar behaviour for a finite interval of time. After this interval of time, the trajectories move away from the stable hyperbolic periodic orbit $\Gamma$ following the unstable manifold $W^u(\Gamma)$. This different behaviour generates the singularities in the Lagrangian descriptors \cite{lopesino2017}.

The variant of Lagrangian descriptor in the equation \eqref{eq:LD} has negative sign between the two integrals. This sign allows differentiating between the stable and unstable manifolds on the Lagrangian descriptor plots. The set of singularities on integral with $\tau_+$ shows the stable manifolds and the set of singularities on integral with $\tau_-$ the unstable manifolds. The superposition of the two patterns of singularities reveals the tangle between the stable and unstable manifolds.

\section{Periodic orbits and their stable and unstable manifolds}
\label{sec:PO}

In order to show the dynamics of this 2 degree of freedom Hamiltonian system in the regime of dissociation and roaming, let us consider a fixed value of the total energy above the dissociation energy threshold, $E = 101 > D_e $. In this case, the trajectories close to the wells can cross the interaction region and escape to infinity. The trajectories that start in the interaction region are sensitive to initial conditions. That means that two trajectories that start arbitrarily close can finish in a different direction in the asymptotic region. This phenomenon is called chaotic scattering and is a kind of transient chaos \cite{Tel_book, Tel2015}. 

The chaotic scattering is a common phenomenon in the open Hamiltonian system. Detailed studies of chaotic scattering in Hamiltonian systems with 2 and 3 degrees of freedom are in the recent literature \cite{Sanjuan2013, Zapfe2010, Drotos2014, Gonzalez2012}. The tangle between the stable and unstable manifolds generates rich dynamics in the interaction region. However, the dynamics is simple in the asymptotic region, where the trajectories converge to straight lines. 

The double Morse potential energy $V$ satisfies the asymptotic condition $Vr^2 \rightarrow 0$ when $r\rightarrow \infty$, which implies that the linear momenta of the trajectories that reach the asymptotic region converge to a constant value \cite{Newton_book}. 

The trajectories associated with dissociation reactions start in one well and finish in the asymptotic region. The trajectories associated with roaming reactions start in one well, travel outside the well in the roaming region where the potential energy is almost flat, and reach the other potential well.

\begin{figure}
\begin{center}
\includegraphics{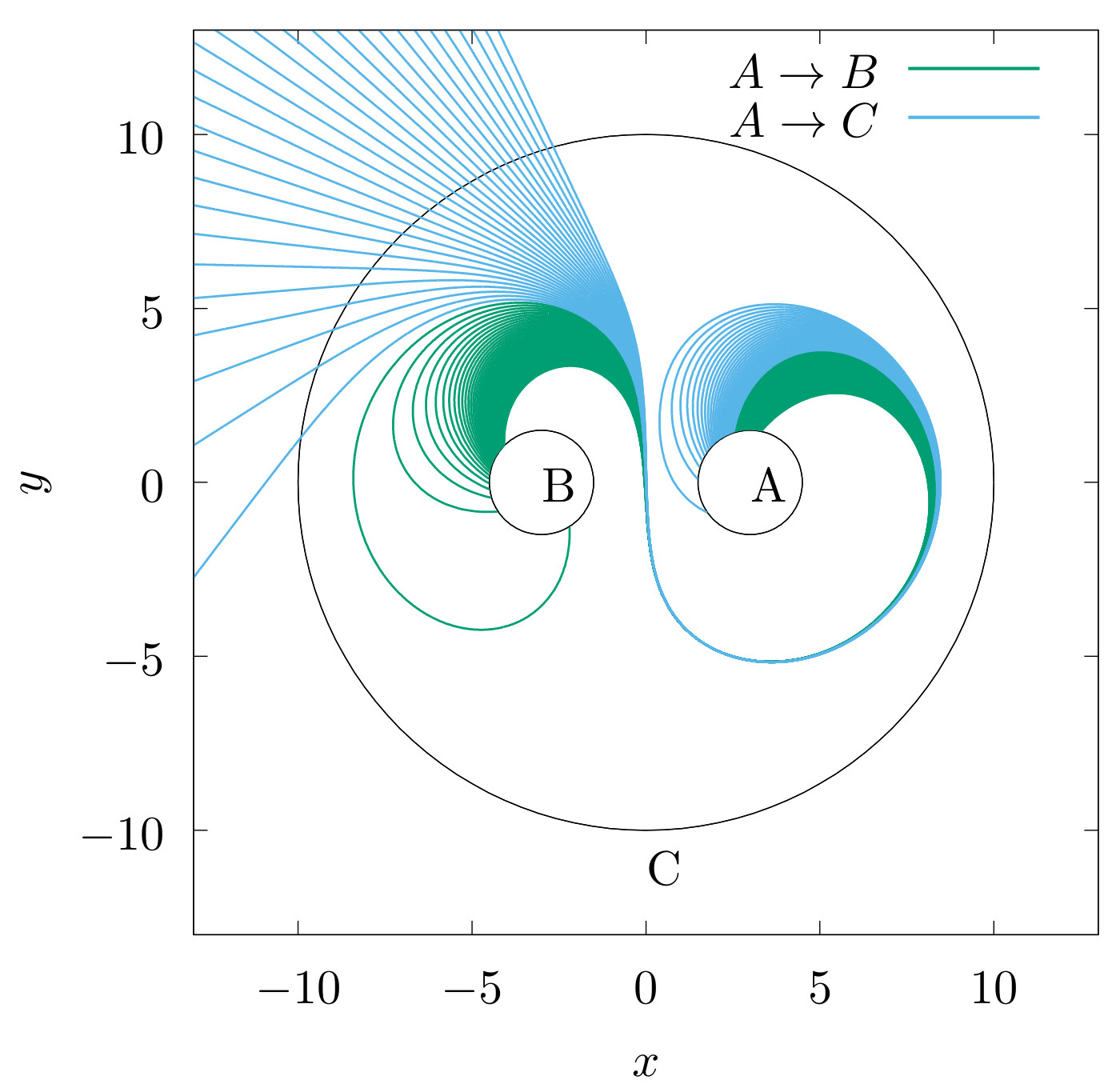}
\caption{ Roaming and dissociation trajectories. The trajectories in this plot start in the region $A$ until intersecting the region $B$ or go to the asymptotic region $C$ defined by an outer circle. The trajectories that travel from one well to the other well are the roaming trajectories, and the trajectories that do not reach the other well are the dissociation trajectories. 
\label{fig:transport}}
\end{center}
\end{figure}

In order to define the transport between the two potential wells and the asymptotic region, the configuration space is divided into three regions, see figure \ref{fig:transport}. Two circumferences around each potential well define the regions $A$ and $B$, and another external circumference around the two wells defines the asymptotic region $C$ where the trajectories are close to straight lines and escape to infinity.

%figure4

% tangles and LD

\begin{figure}
\begin{center}
\includegraphics{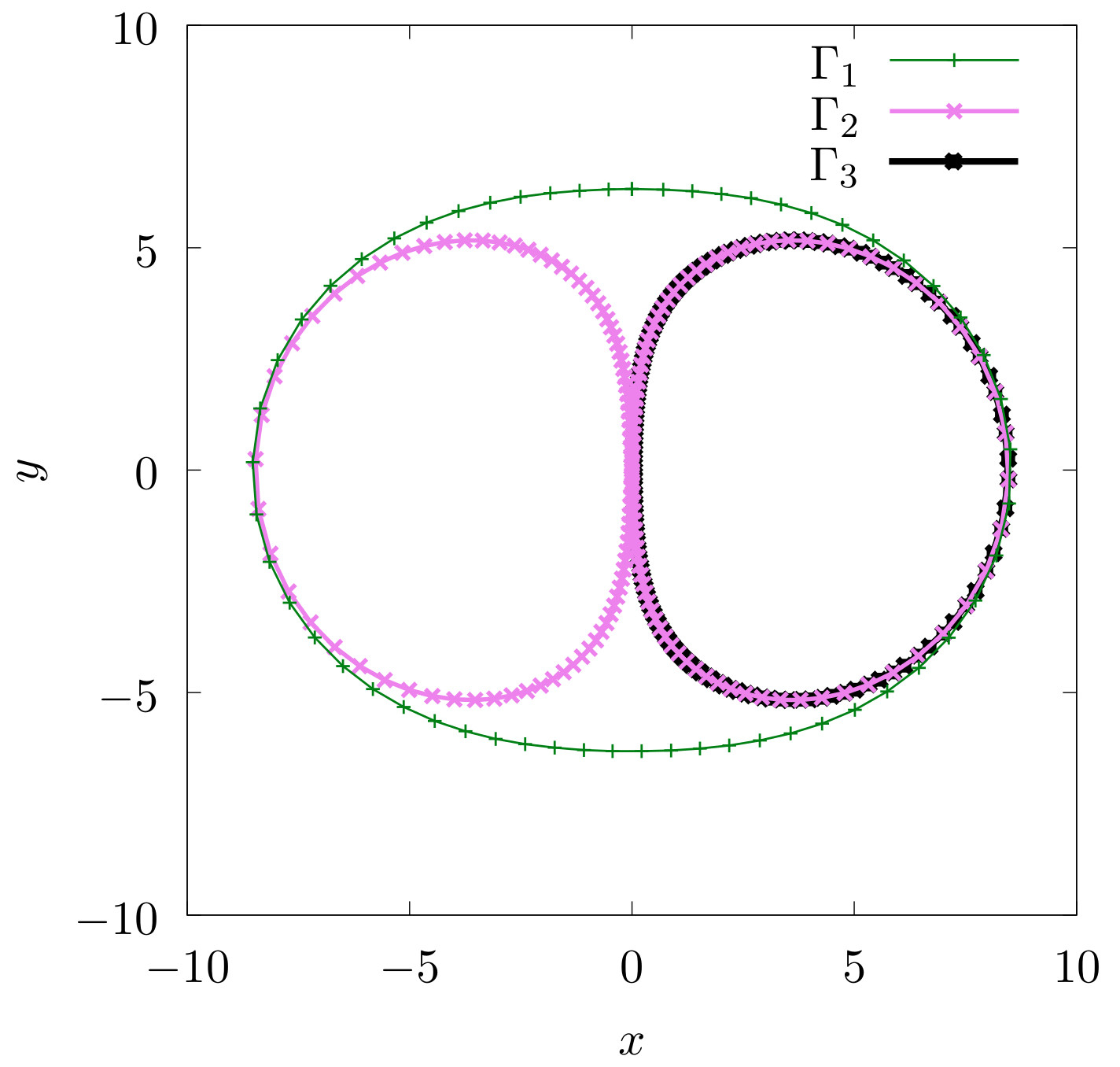}
\caption{Projections of the periodic orbits $\Gamma_1$, $\Gamma_2$, and $\Gamma_3$ in the configuration space.  
\label{fig:orbits_2D}}
\end{center}
\end{figure}

A relevant conclusion in the work \cite{carpenter2018dynamics}, is that the roaming and dissociation trajectories in the system are related to three types of periodic orbits. Let be $\Gamma_1$, $\Gamma_2$, and $\Gamma_3$ periodic orbits representatives of each type. The projections of the three periodic orbits in configuration space is shown in figure \ref{fig:orbits_2D}. The projection of periodic orbit $\Gamma_1$ encircles the interaction region and the two potential wells. The projection of $\Gamma_2$ also encircles the two wells but self intersects at the origin. The projection of $\Gamma_3$ encircles just one well.

\begin{figure}
\begin{center}
\includegraphics{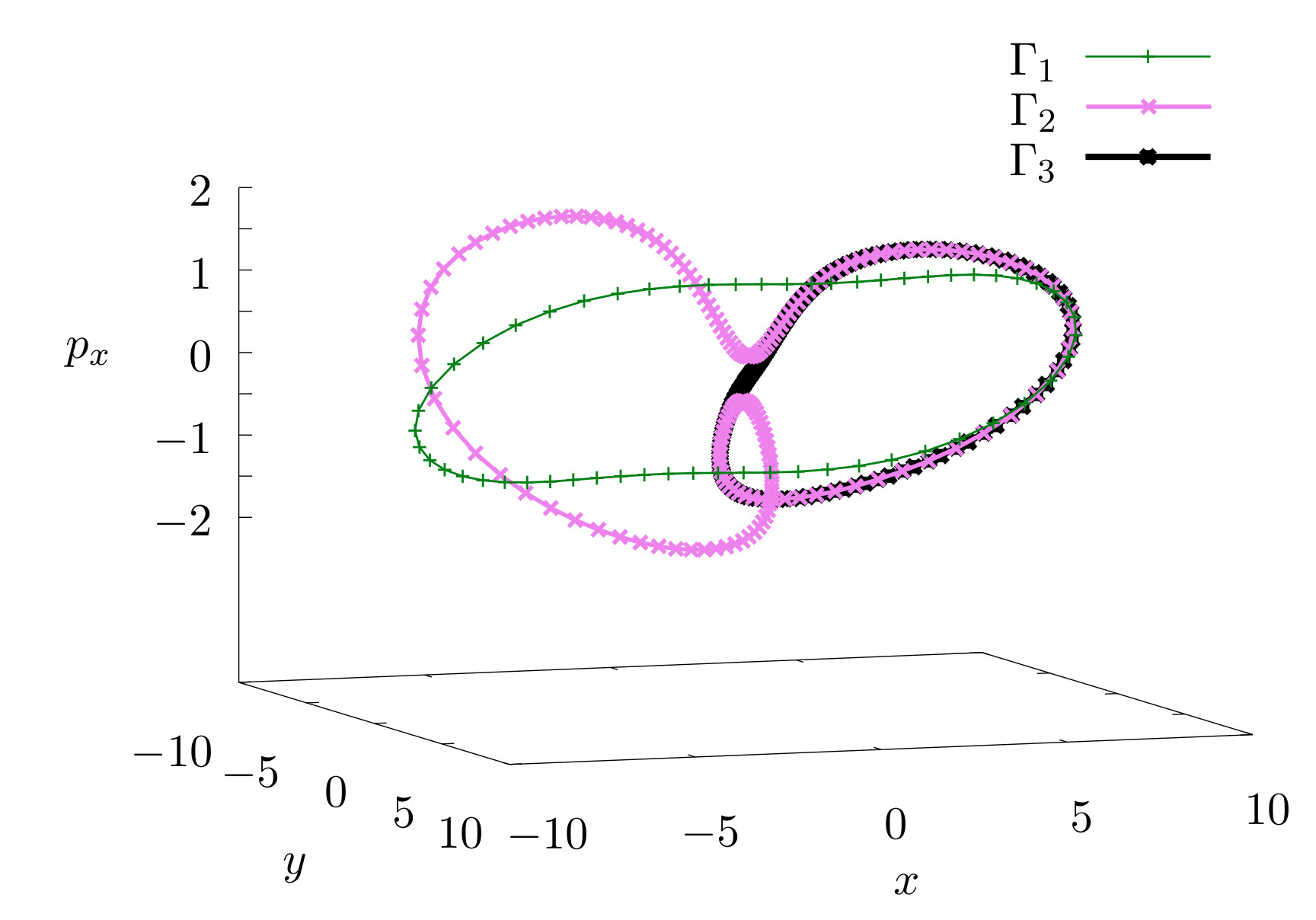}
\caption{ Periodic orbits $\Gamma_1$, $\Gamma_2$, and $\Gamma_3$ in the constant energy manifold parameterised by coordinates $(x,y,p_x)$. 
\label{fig:orbits_3D}}
\end{center}
\end{figure}

In the configuration space, the projection of periodic orbits $\Gamma_2$, $\Gamma_3$ are close, but it is clear that the orbits are separated in the constant energy manifold, see figure \ref{fig:orbits_3D}. The three orbits are very close to each other in a neighbourhood of the point $(x=8.45,y=0,p_x=0)$.

As a consequence of the symmetries $x\rightarrow -x$ and $y \rightarrow -y$ of the potential energy $V(x,y)$, there are 4 orbits of type I, 2 orbits of type II, and 4 orbits of type III in the phase space. It is necessary to consider only one element of every type for the numerical calculations. 

%figure5

%figure6

A common method to find the periodic orbits is based on the Poincar\'e map defined on a 2-dimensional Poincar\'e surface. The Poincar\'e map is a function from the Poincar\'e surface to itself obtained by following trajectories from one intersection of the Poincar\'e surface to the next. The fixed points of the Poincar\'e map are the intersections between the periodic orbits and the Poincar\'e surface. In this system, a natural choice of Poincar\'e surface is the canonical plane $x$--$p_x$ with $y=0$ due to the symmetry $y\rightarrow -y$ of the potential energy. The three periodic orbits $\Gamma_1$, $\Gamma_2$, and $\Gamma_3$ are hyperbolic and considerably unstable. In order to calculate the eigenvalues and eigenvectors of the linear approximation to the Poincar\'e map around the fixed points associated with the three types of periodic orbits is necessary to use multi-precision integrators. These calculations are done with the Taylor integrator method implemented in Julia \cite{Perez2019}. The periods of the periodic orbits and their associated eigenvalues are in Table \ref{table}. The product between the eigenvalues is close to 1, so the calculations agree with the Hamiltonian properties of the system.

\begin{table}

\begin{center}
\begin{tabular}{ | l | l | l | l | l |}
\hline
$\Gamma_1$ & $T_{1} =  30.3774  $ & $\lambda_{1,Max}\simeq 2.0314 \times 10^7$ & $\lambda_{1,min}\simeq 4.9224 \times 10^{-8}$ & $\lambda_{1,Max} \lambda_{1,min} \simeq 0.99999 $  \\
%\hline
$\Gamma_2$ & $T_{2} = 29.1727$ & $\lambda_{2,Max}\simeq 3.4627 \times 10^{12}$ & $\lambda_{2,min}\simeq 2.8965 \times  10^{-13}$ & $\lambda_{2,Max} \lambda_{2,min} \simeq 1.00299 $  \\
%\hline
$\Gamma_3$ & $T_{3} =  14.5671  $ & $\lambda_{3,Max}\simeq 1.8019 \times 10^6$ & $\lambda_{3,min} \simeq 5.5382 \times 10^{-7}$ &  $\lambda_{3,Max} \lambda_{3,min} \simeq 0.99794 $  \\
\hline

\end{tabular}
\end{center}
\caption{Periods and eigenvalues of the periodic orbits $\Gamma_1$, $\Gamma_2$, and $\Gamma_3$. \label{table}}

\end{table}

It is possible to construct the stable and unstable manifolds of the periodic orbits using their corresponding eigenvalues and eigenvectors. However, the construction and visualisation of the stable and unstable manifolds are not simple tasks because of the size of the eigenvalues, $\lambda_{k,Max} >> 1$ and $\lambda_{k,min} \cong 0$ for $k =1,2,3$, and because the Poincar\'e map requires that the trajectories intersect the Poincar\'e section again. For example, the trajectories in the unstable manifold of the periodic orbit $\Gamma_1$ that go directly to the asymptotic region intersect the Poincar\'e section only for a finite number of iterations before to converging to straight lines. In general, it is not possible to obtain a global visualisation of the tangles between the stable and unstable manifolds using the Poincar\'e map. 

An appropriate approach that allows an easier visualisation of the stable and unstable manifolds even in this situation is the use of indicators for chaotic regions, like fast Lyapunov indicators (FLI) \cite{Lega2016}, mean exponential growth factor of nearby orbits (MEGNO) \cite{Cicotta2016}, the smaller (SALI) and the generalized (GALI) alignment indices \cite{Skokos2016}, delay time functions, scattering functions \cite{Gonzalez2012}, and Lagrangian descriptors. The Lagrangian descriptors contain information about the stable and unstable manifolds that intersect the set of initial conditions and do not require that the trajectories intersect any particular predefined set again, as mentioned in section \ref{sec:LD}. It is important to notice that the trajectories should be integrated for long enough time to reach a neighbourhood of the geometric structures that generate different abrupt behaviour. In the other case, the chaotic indicator does not show the structure clearly, even if the set of initial conditions intersect the structure.
    
The calculations of the Lagrangian descriptors are done with the classical ninth order Vernet integrator implemented in Julia \cite{Rackauckas2017}. In this system, the Lagrangian descriptors require standard double-precision to reveal segments of the invariant manifolds. The computational cost necessary to see a signature of the stable and unstable manifolds in the Lagrangian descriptor plots is a fraction of the computational cost necessary using the Poincar\'e map. The reason for this difference is related to the computational cost of the integration with the multi-precision libraries and the fact that the calculation of the Lagrangian descriptor does not require that the trajectories cross the set of initial conditions again.

\begin{figure}
 \begin{center}
 
 \subfigure[$\tau_{+}=6,\tau_{-}=-6$]{  
 %\label{}
 \scalebox{1}{
 \includegraphics{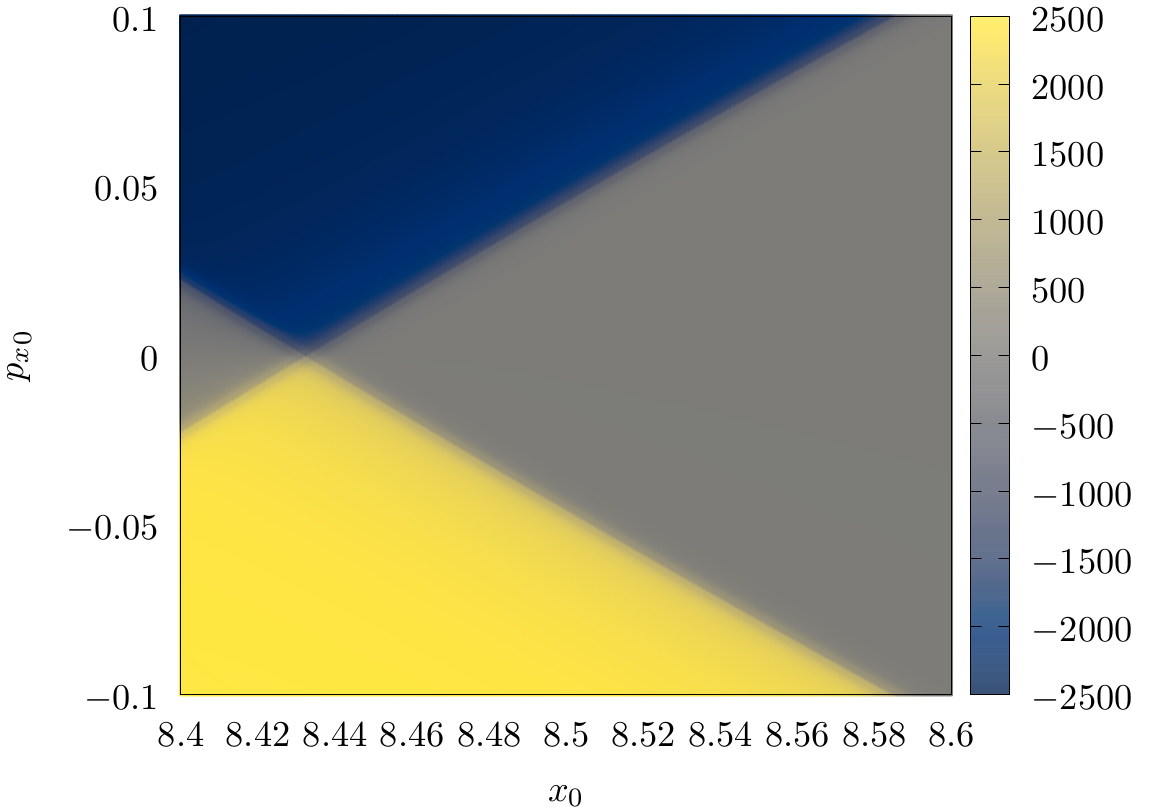}  }}
% &
 \subfigure[$\tau_{+}=8,\tau_{-}=-8$ ]{
 %\label{}
 \scalebox{1}{
 \includegraphics{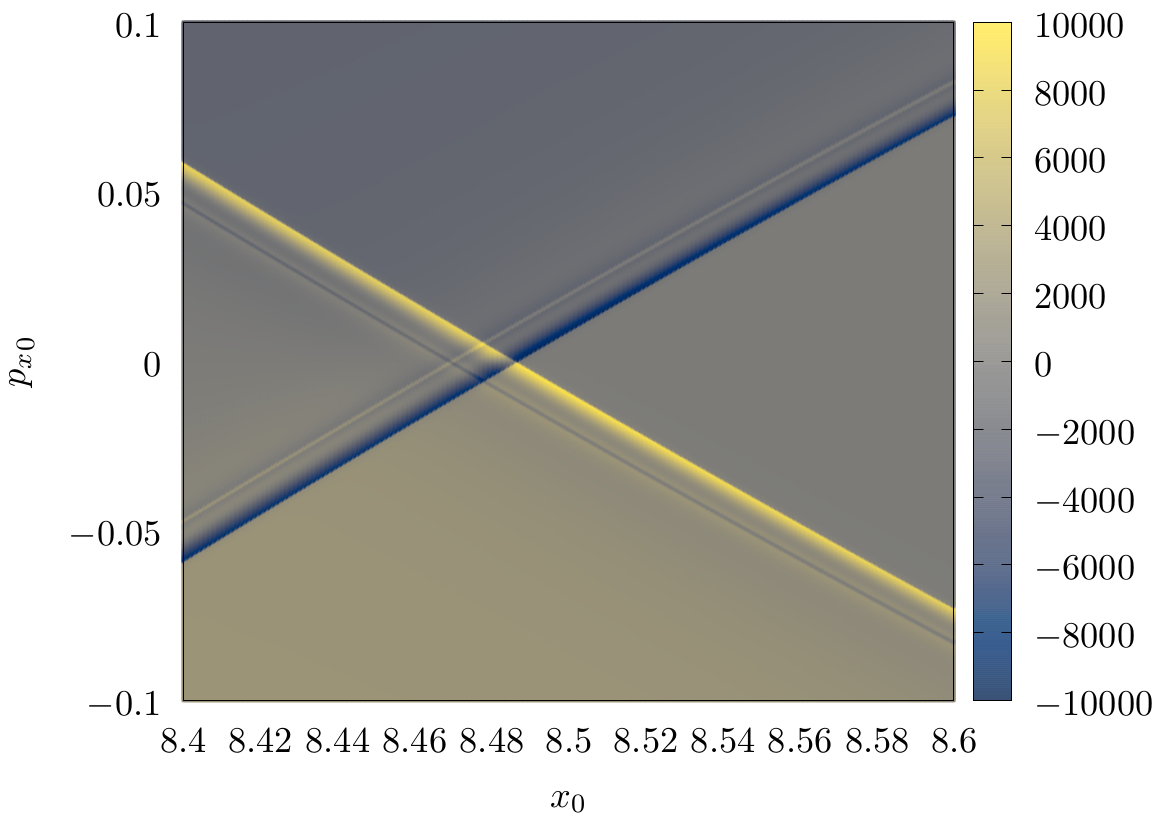}  }}\\
% &
\subfigure[$\tau_{+}=10,\tau_{-}=-10$]{
 %\label{}
 \scalebox{1}{
 \includegraphics{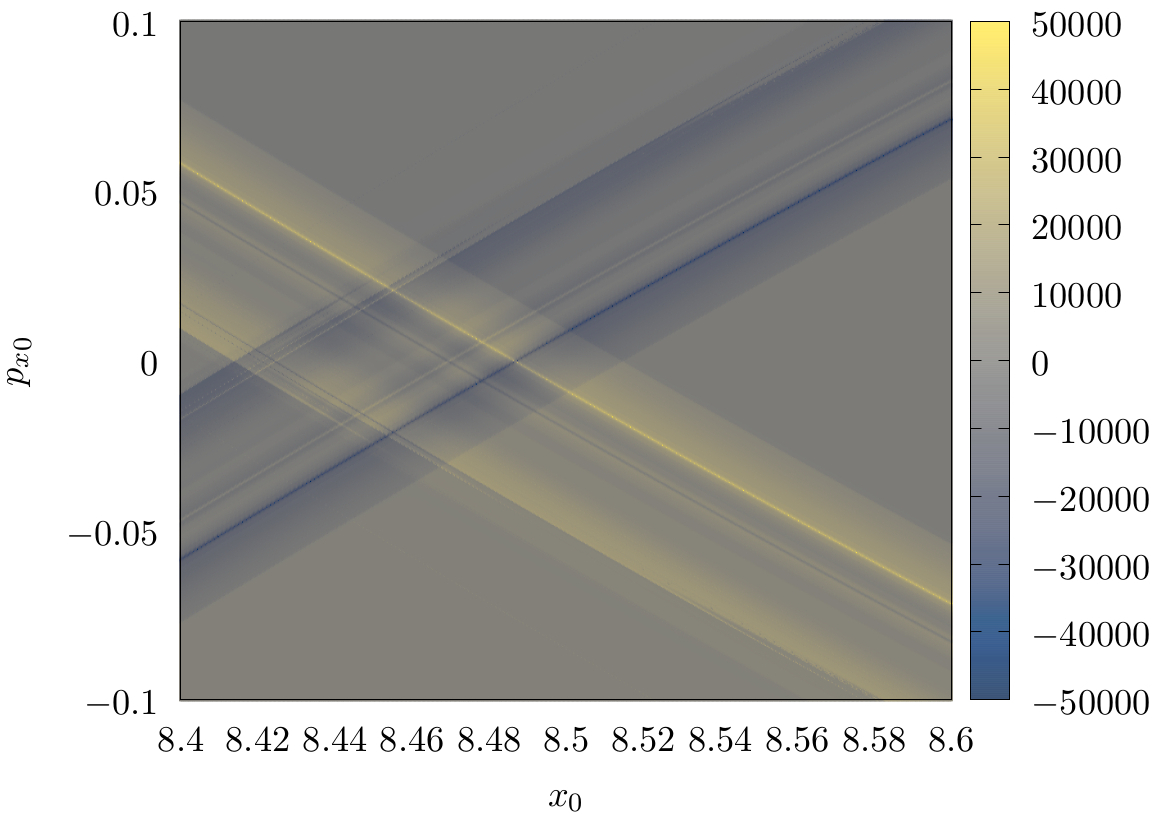}  }}
% & 
\subfigure[$\tau_{+}=12,\tau_{-}=-12$]{
% \label{}
 \scalebox{1}{
 \includegraphics{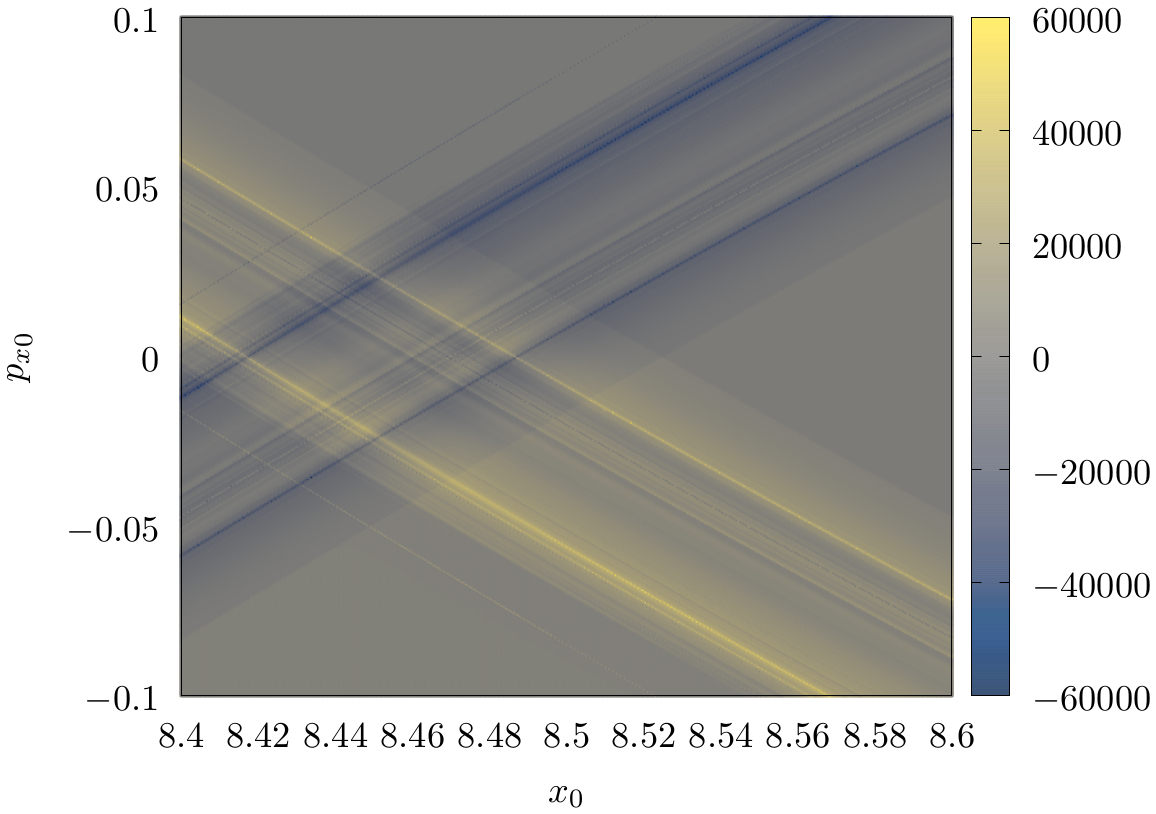} }} \\
\subfigure[$\tau_{+}=15,\tau_{-}=-15$]{  
% \label{}
 \scalebox{1}{
 \includegraphics{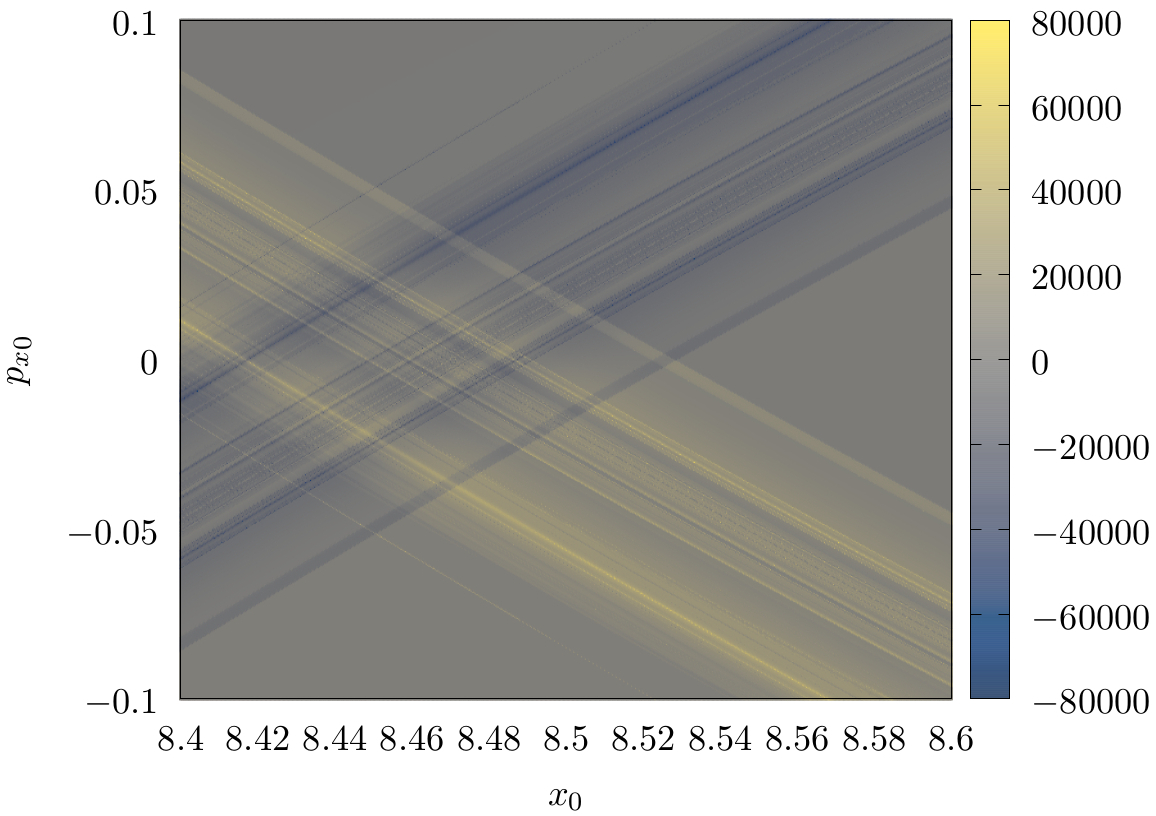}  }}
% &
 \subfigure[$\tau_{+}=17,\tau_{-}=-17$ ]{
% \label{}
 \scalebox{1}{
 \includegraphics{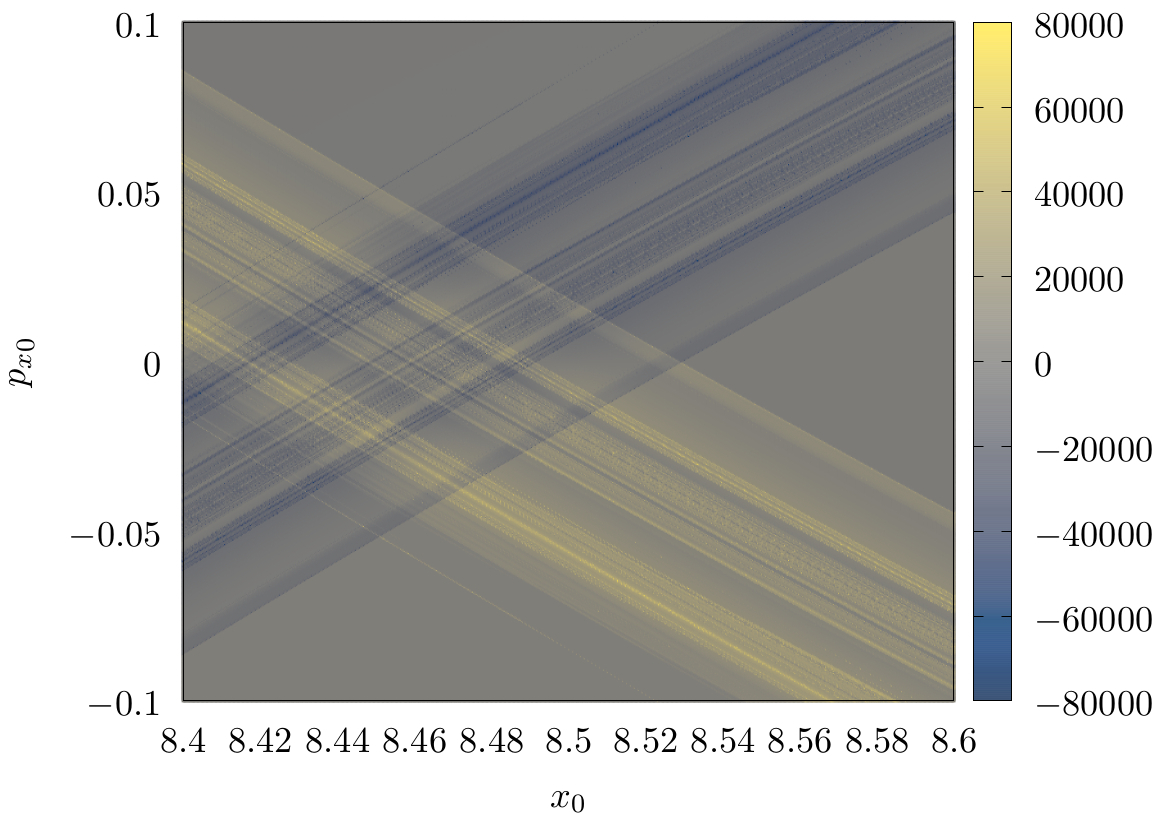}  }}

\caption{Lagrangian descriptor $M$ evaluated on the plane $p_{x_0}$--$x_0$ with initial $y_0 = 0$ and initial momentum $(p_{y_0} < 0, p_x = 0)$ for different values of $\tau_+$ and $\tau_-$. To show the symmetry of $M$ and the invariant manifolds for this initial conditions, $\tau_+ = - \tau_-$ is selected. When the value of $\tau_+$ is increased, the structure of the tangle between the stable and unstable manifolds is revealed by the abrupt jumps in the value of the Lagrangian descriptor where it converges to a non-differentiable function. The intense blue and yellow curves correspond to segments of the unstable manifolds and stable manifolds respectively. 
\label{fig:LD_time}}
\end{center}
\end{figure}

The figure \ref{fig:orbits_2D} shows that $\Gamma_1$, $\Gamma_2$, and $\Gamma_3$ intersect the $x$ axis in a neighbourhood of the point $(x= 8.5,y=0)$ and their momentum at the intersection are $(p_{y} < 0, p_x = 0)$. In order to find these periodic orbits it is natural to calculate the Lagrangian descriptor evaluated on the canonical conjugate plane $p_{x_0}$--$x_0$, $y_0 = 0$, and initial momentum $ p_{y_0} < 0$. The Lagrangian descriptor evaluated in this set of initial conditions for different times is shown in figure\ref{fig:LD_time}. When the size of the interval of integration $[\tau_-,\tau_+]$ is increased, the Lagrangian descriptor reveals intersection of the the stable and unstable manifolds with the set of initial conditions. The intersections between the stable and unstable manifolds is the invariant chaotic set. The periodic orbits $\Gamma_1$, $\Gamma_2$, and $\Gamma_3$ are contained in this invariant chaotic set.

%figure7

\begin{figure}
 \begin{center}
 
 \subfigure[$\Gamma_1$]{  
 \label{fig:LD_line_Gamma1}
 \scalebox{0.4}{
 \includegraphics{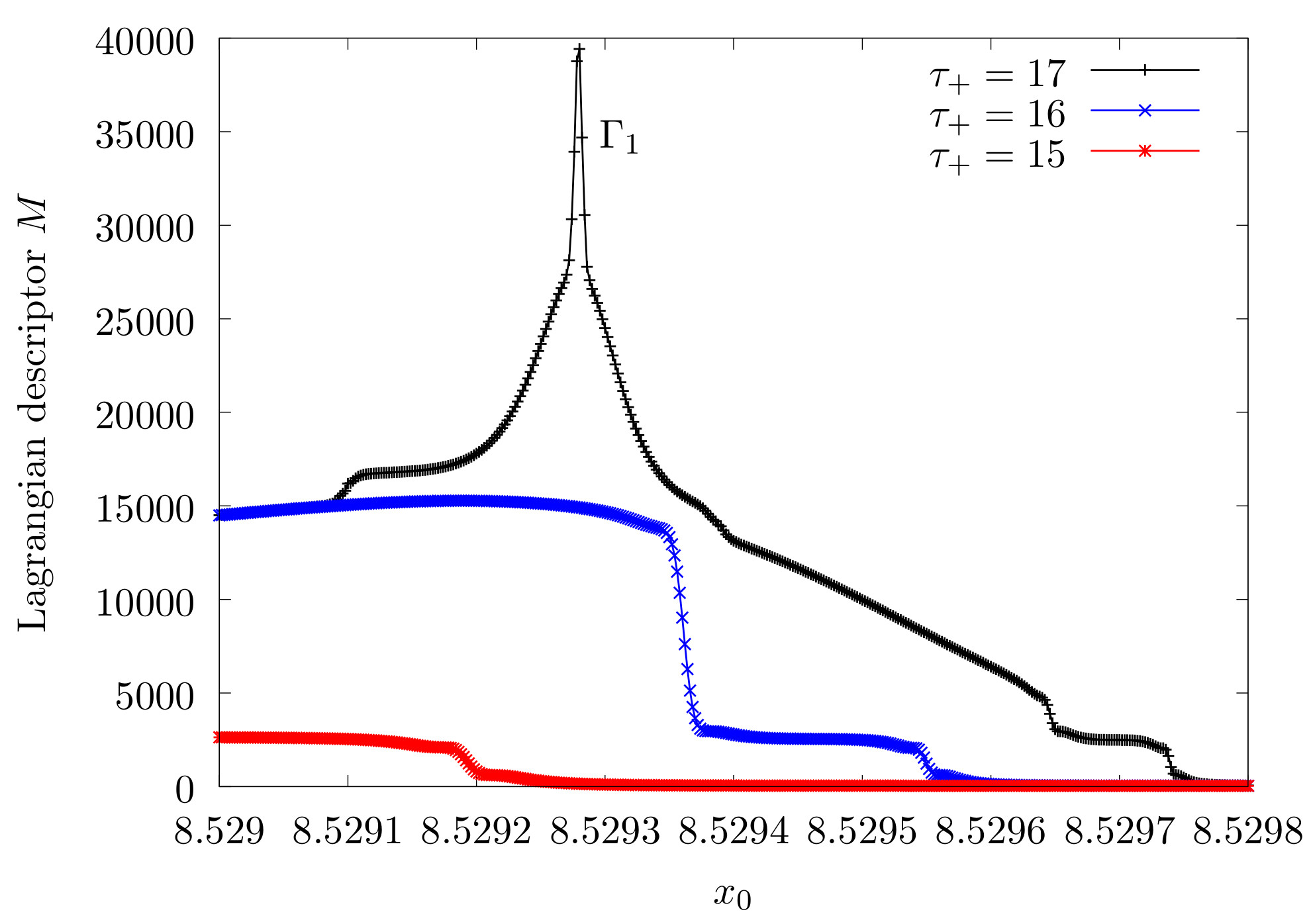}  }}
 \subfigure[$\Gamma_2$ ]{
 \label{fig:LD_line_Gamma2}
 \scalebox{0.4}{
 \includegraphics{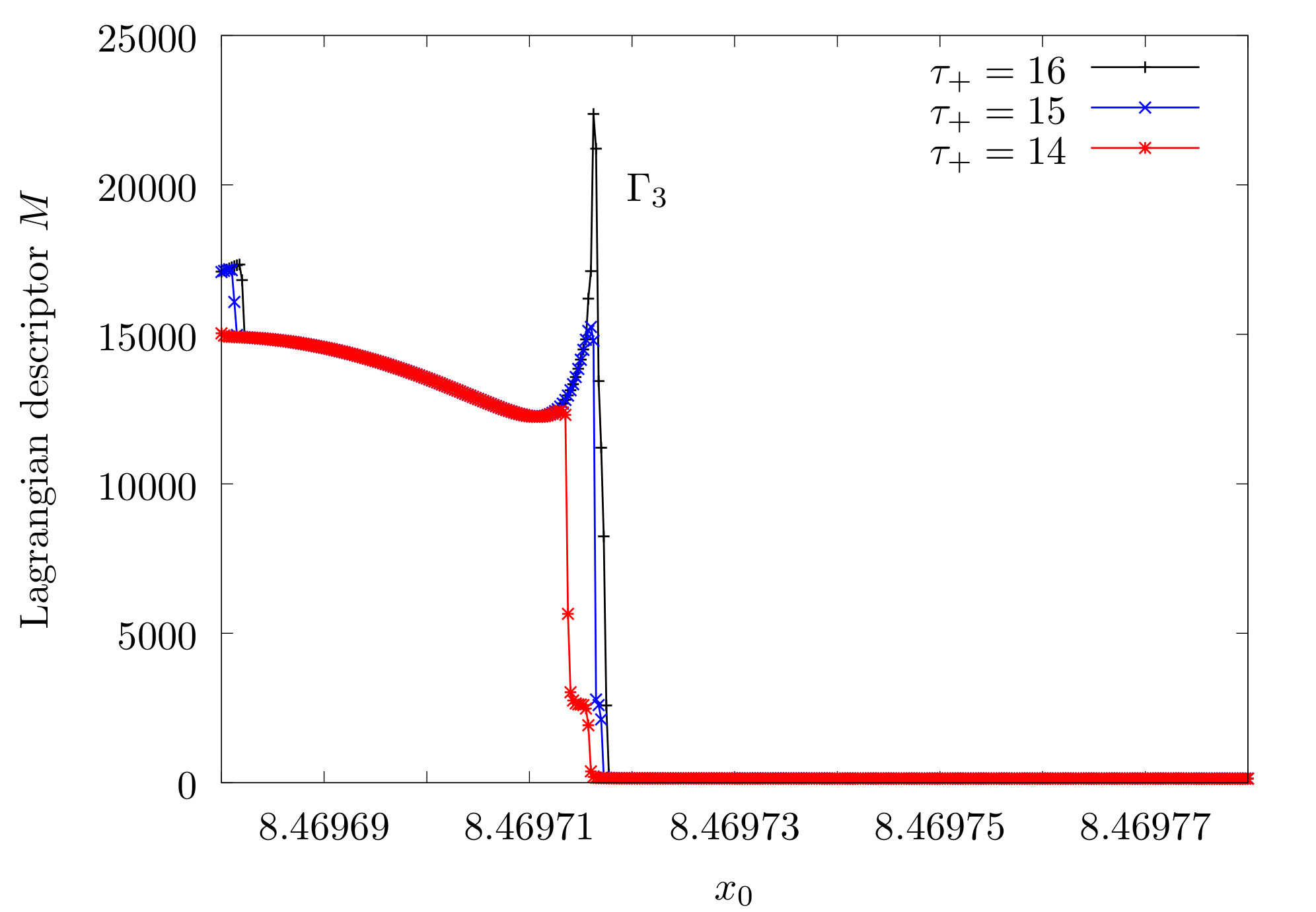}  }}
 \subfigure[ $\Gamma_3$]{  
 \label{fig:LD_line_Gamma3}
 \scalebox{0.4}{
 \includegraphics{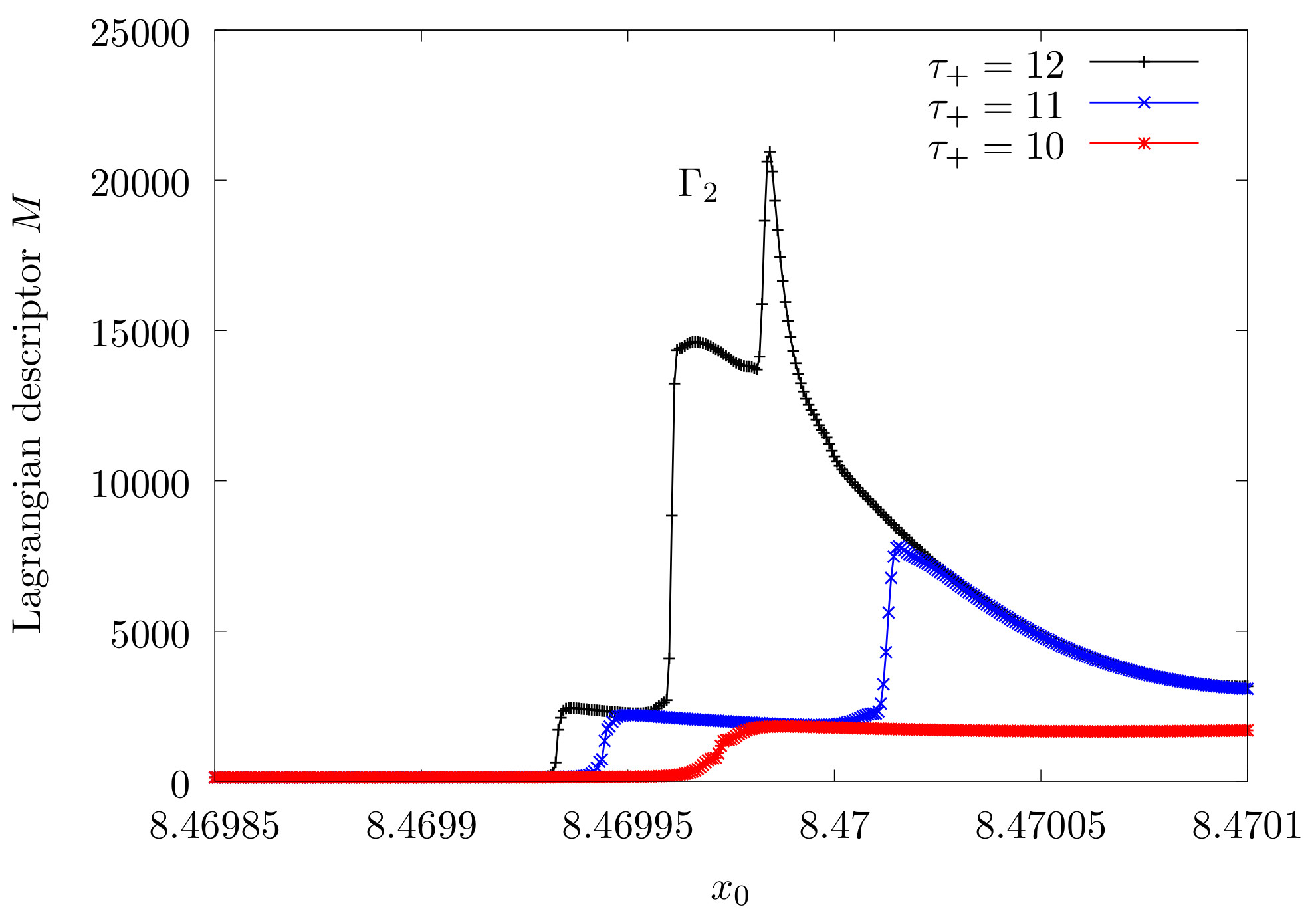}  }}\\

\caption{ Lagrangian descriptors evaluated on different intervals of the line $x_0$. For all the plots, the integration times are $\tau_- = 0$ and $ \tau_+ > 0$. The different colour curves in the plots correspond to different values of $ \tau_+$. The peaks indicate the intersection of the periodic orbits $\Gamma_1$, $\Gamma_2$, and $\Gamma_3$ with the line $x_0$.  
\label{fig:LD_line_x0}}
\end{center}
\end{figure}

In order to show clearly the peaks generated by invariant stable and unstable manifolds of the three hyperbolic periodic orbits in the Lagrangian descriptor plots and find the periodic orbits, let us consider the symmetry line $y_0 = 0$ in intervals that contain the periodic orbits. The Lagrangian descriptor in figure \ref{fig:LD_time} with $\tau_{-} = -\tau_+$ evaluated on this line is identically zero due to the symmetries of the system. For this reason, is convenient to consider the Lagrangian descriptor with $\tau_{-} = 0$ and $\tau_+ > 0$, the results are in the figure \ref{fig:LD_line_x0}. When the integration time $\tau_+$ is increased enough, the peak generated for the different behaviour of the trajectories reveals the intersections of the stable manifold of the periodic orbit with the $x_0$ axis. Due to the symmetry of the stable and unstable manifolds with respect to the $x_0$ axis, this intersection coincides with the intersection with the unstable manifold of the periodic orbit. Finally, the intersection point between the stable and unstable manifolds is the intersection with the periodic orbit with the $x_0$ axis. It is important to notice that the integration time to generate the peak is different from the period of the orbit. For the periodic orbits $\Gamma_1$ and $\Gamma_2$ the time necessary to generate the peak is less than their corresponding periods.

Another simple way to find the intersections between the stable and unstable manifolds is to change the negative sign in the definition \eqref{eq:LD}, then $M_+ + M_- > 0 $ and the points where is this quantity is maximal are the intersections between the stable and unstable manifolds.

\begin{figure}
\begin{center}%
\includegraphics{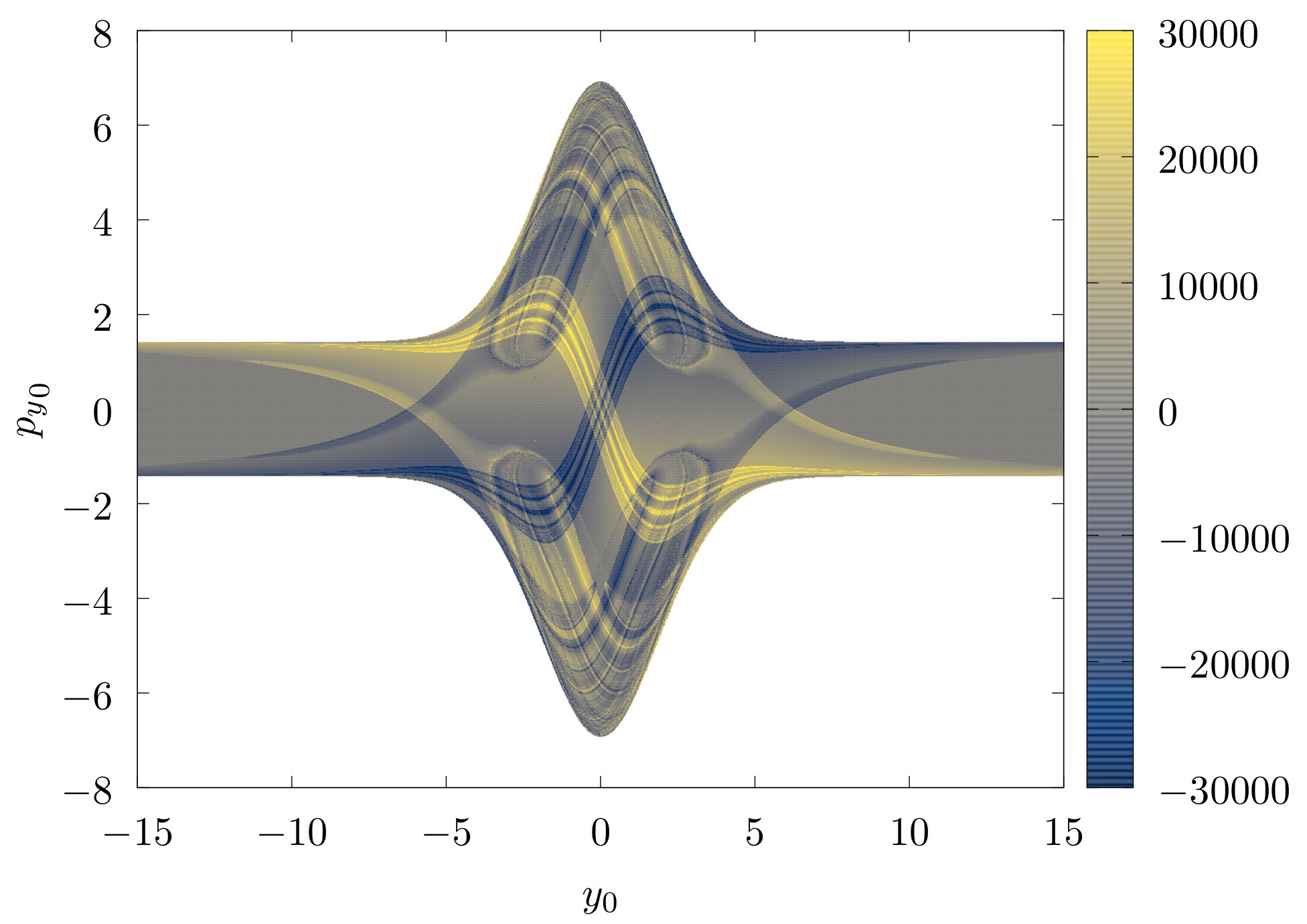}
\caption{Lagrangian descriptor $M$ evaluated on the plane $p_{y_0}$--$y_0$ at $x_0=0$, with initial momentum $p_{x_0}>0$.
\label{fig:LD_plane_x=0}}
\end{center}
\end{figure}

\begin{figure}
\begin{center}%
\includegraphics{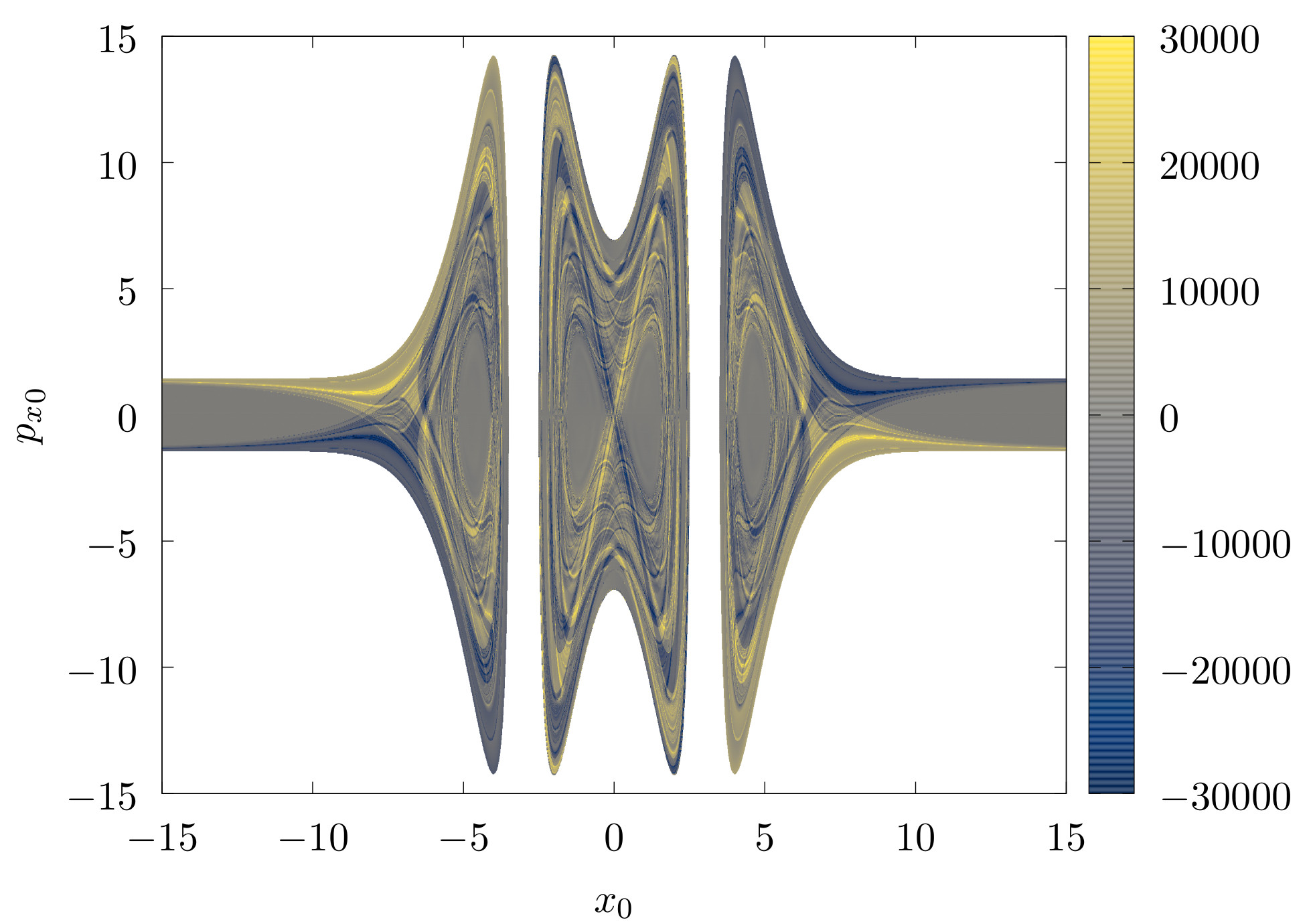}
\caption{Lagrangian descriptor $M$ evaluated on the plane $p_{x_0}$--$x_0$ at $y_0=0$ with initial $p_{y_0}<0$.
\label{fig:LD_plane_y=0}}
\end{center}
\end{figure}

The structure of the tangle is very rich in the interaction region. To see this structure let us consider the Lagrangian descriptor evaluated on large areas of the canonical conjugate planes $p_{y_0}$--$y_0$ at $x_0=0$ with     $p_{x_0}>0$, and $p_{x_0}$--$x_0$ at $y_0=0$ with initial $p_{y_0}<0$. The Lagrangian descriptors on the figures \ref{fig:LD_plane_x=0} and \ref{fig:LD_plane_y=0} show the symmetries of the tangles and their complicated structure characteristic of the open chaotic Hamiltonian systems with 2 degrees of freedom. In both figures, some lines converge directly to constant values of the momentum in the asymptotic region. These lines correspond to the branches of stable and unstable manifolds that go directly to the asymptotic region. Close to these branches there are the lobules where the particles escape or enter to the interaction region. Examples of the lobule dynamics in open Hamiltonian systems and chaotic scattering are in \cite{Gonzalez2012,Zapfe2010}.

In figure \ref{fig:LD_plane_y=0} there are three disconnected regions: two unbounded regions on the extremes and one bounded region in the middle surrounded by a white region. The white region corresponds to points where the momentum $p_{x_0}$ is not defined for this value of the energy.
    
\newpage

\section{Phase space pathways for roaming}
\label{sec:roaming}

% Dividing surface

An important set in the phase space for analysing the transport in the system is the dividing surface of a periodic orbit\cite{Pechukas1979,carpenter2018dynamics}. The projection of the periodic orbit in the configuration space defines a closed curve and a region inside the curve. The dividing surface is a natural surface to analyse all the trajectories of the system that cross the boundary of this region in the configuration space. To construct the dividing surface in the phase space, let us consider all trajectories that cross the boundary of this region in the configuration space. For every point in the boundary, we consider all the possible momenta compatible with the energy of the system. This procedure generates a set of points in the phase space for every point on the boundary of the region in the configuration space. The union of those sets of points forms a surface in the phase space. 

In more detail, the algorithm to construct the dividing surface of the periodic orbit $\Gamma$ consists of three steps: 

\begin{itemize}
  
\item Project the periodic orbit $\Gamma$ on the configuration space.

\item Construct the circumference on momentum plane using the equation 

\[  p^2_x + p^2_y  = 2m(E - V(x,y)) \]

for every point $(x,y)$ on the projection of the orbit $\Gamma$ in the configuration space.

\item Take the union of all these circumferences in the phase space to construct the dividing surface.

\end{itemize}    

The dividing surface constructed in this way has three relevant properties for analysis of the transport: The first property is that the periodic orbit $\Gamma$ and
the orbit with the same projection in configuration space but opposite momentum are contained in the same dividing surface. This fact is a consequence of the time-reversal invariance of the system. 
The second property is that the periodic orbits used to construct the dividing surface are the boundaries between the regions where the trajectories enter into the phase space region contained by the dividing surface and trajectories that left the same region. The third property, the flux through the dividing surface is minimal. That means if the dividing surface is deformed, the flux through it increases. The demonstrations of these properties of the dividing surfaces of periodic orbits are in \cite{Pechukas1979}, and for NHIMs with more dimensions in \cite{Waalkens_2004}. 

In the system, the periodic orbits $\Gamma_1$, $\Gamma_2$, and $\Gamma_3$ are not generated by any saddle in the potential energy. Their projections in the configuration space are curves that enclose an area. In the case of unstable periodic orbits associated with saddles points, the projection on the configuration space is a curve that does not enclose any area. The extremes of the projection are the returning points where the periodic orbit has zero momentum. The dividing surface, in this case, is a surface genus 0 (topologically equivalent to a sphere). An example of this kind of periodic orbit and its corresponding dividing surface is in \cite{Rafa2019}.

The projection in configuration space of the orbits $\Gamma_1$, $\Gamma_3$ do not have self-intersections; for this reason, the corresponding dividing surfaces are genus 1 (a torus). Another example of this kind of dividing surface is in \cite{Collins_2015}. The projection of the periodic orbit $\Gamma_2$ has an intersection at the origin, the dividing surface, in this case, is a surface genus 2. 

A natural parametrisation of the dividing surface is given by the arc length $l_0$ of the respective periodic orbit in the phase space and the angle between the initial momentum and the $x_0$ axis, $\phi_0 = \displaystyle{ \arctan{(p_{y_0}/p_{x_0}})}$. The next calculation uses this parametrisation
to analyse the trajectories that cross each of the three dividing surfaces.

%% Fate maps on DS

\begin{figure}
 \begin{center}
 
 \subfigure[$ $]{  
 \label{fig:FM_DS_Gamma1}
 \scalebox{0.6}{
 \includegraphics{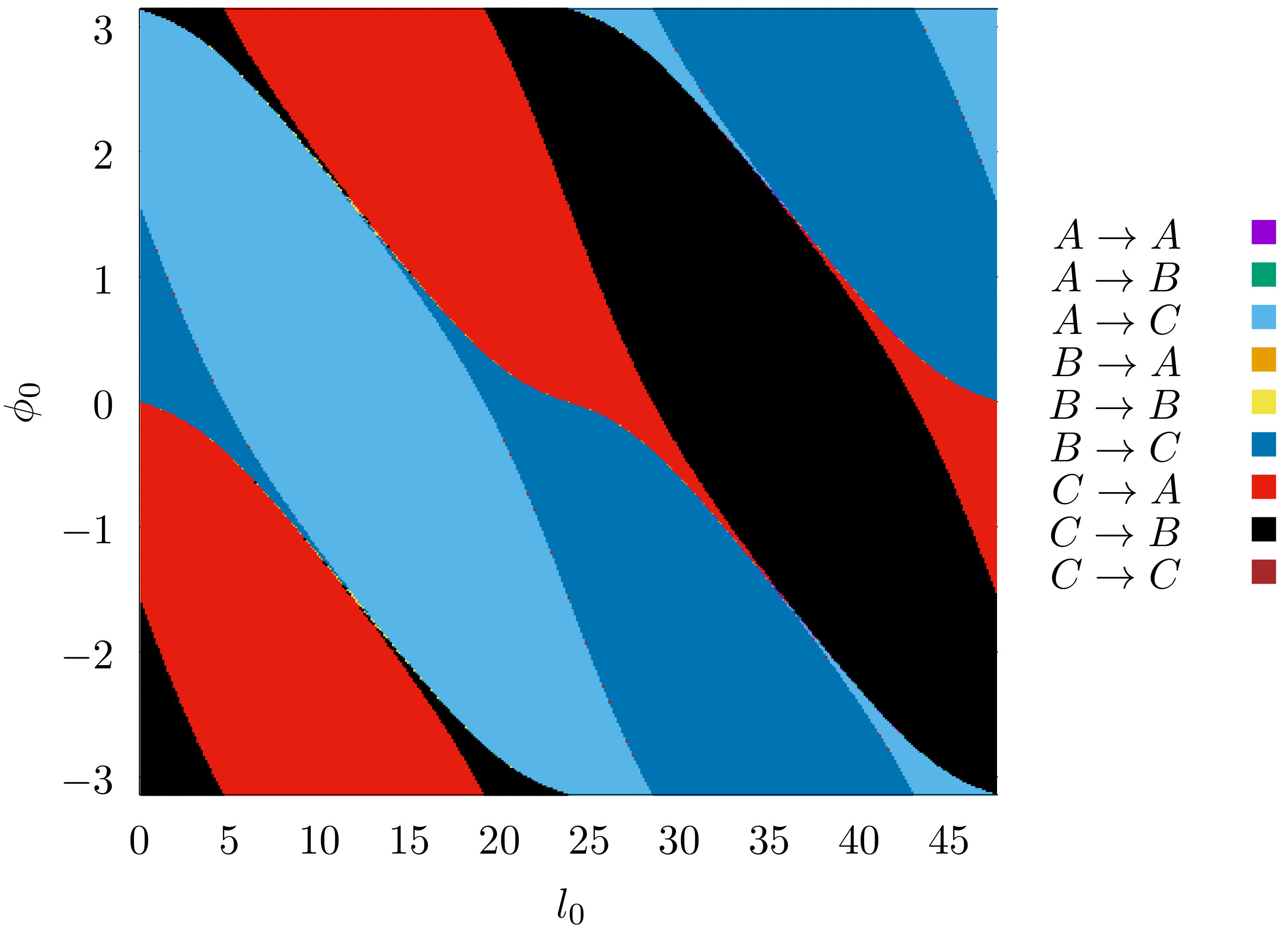}  }}
% &
 \subfigure[$ $ ]{
 \label{fig:LD_DS_Gamma1}
 \scalebox{0.6}{
 \includegraphics{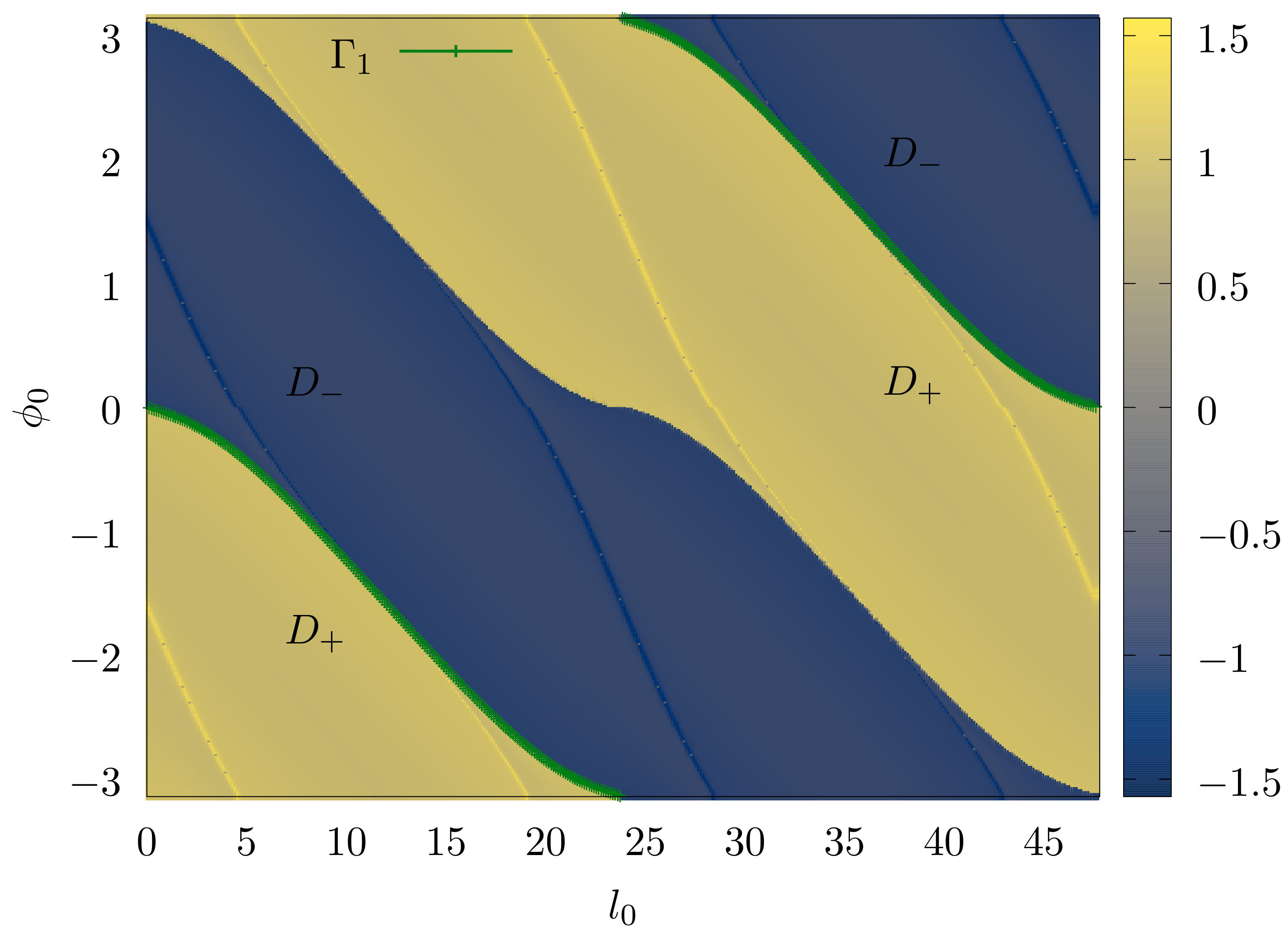}  }}\\
  \subfigure[$ $]{  
 \label{fig:FM_DS_Gamma2}
 \scalebox{0.6}{
 \includegraphics{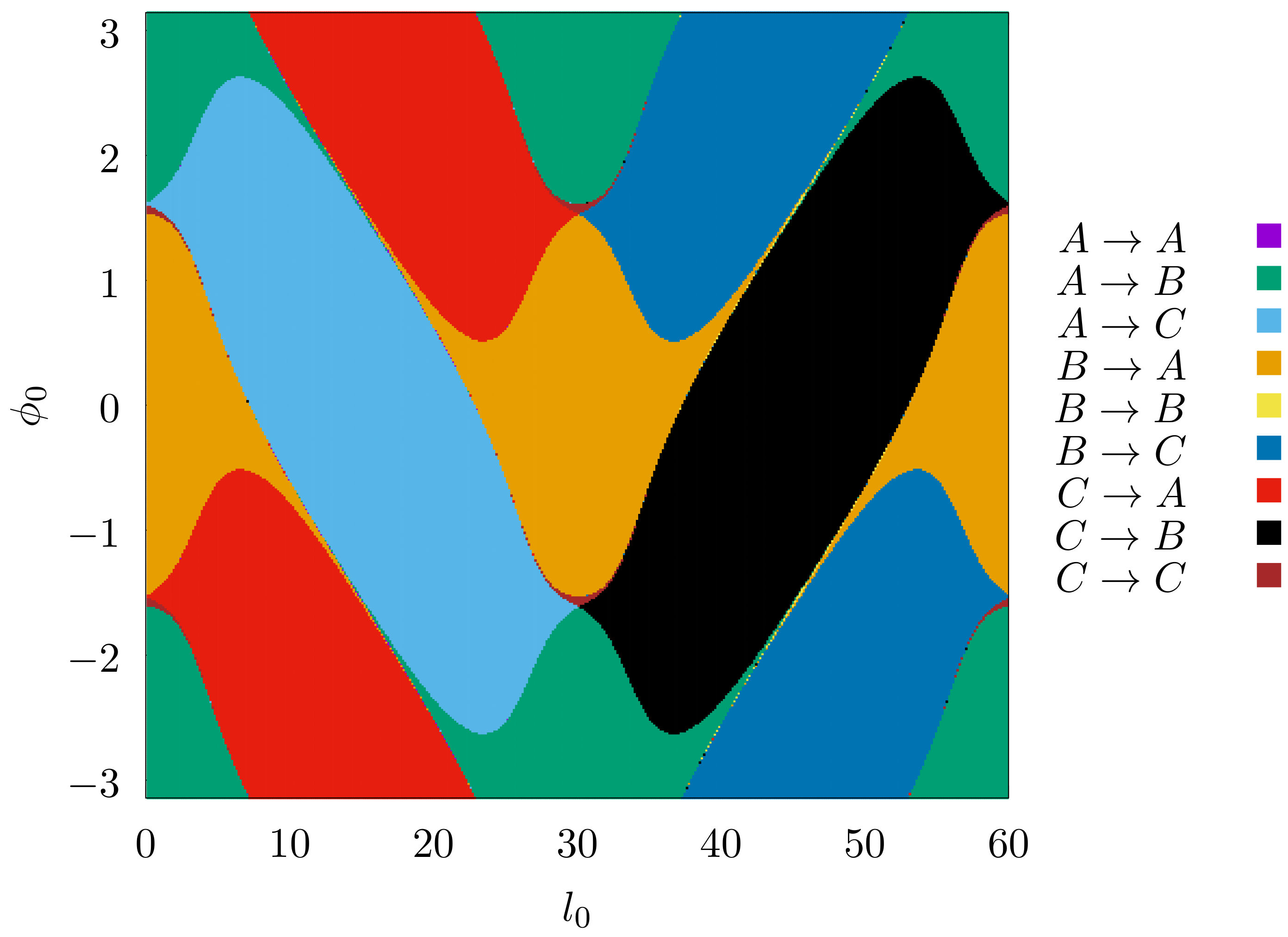}  }}
% &
 \subfigure[$ $ ]{
 \label{fig:LD_DS_Gamma2}
 \scalebox{0.6}{
 \includegraphics{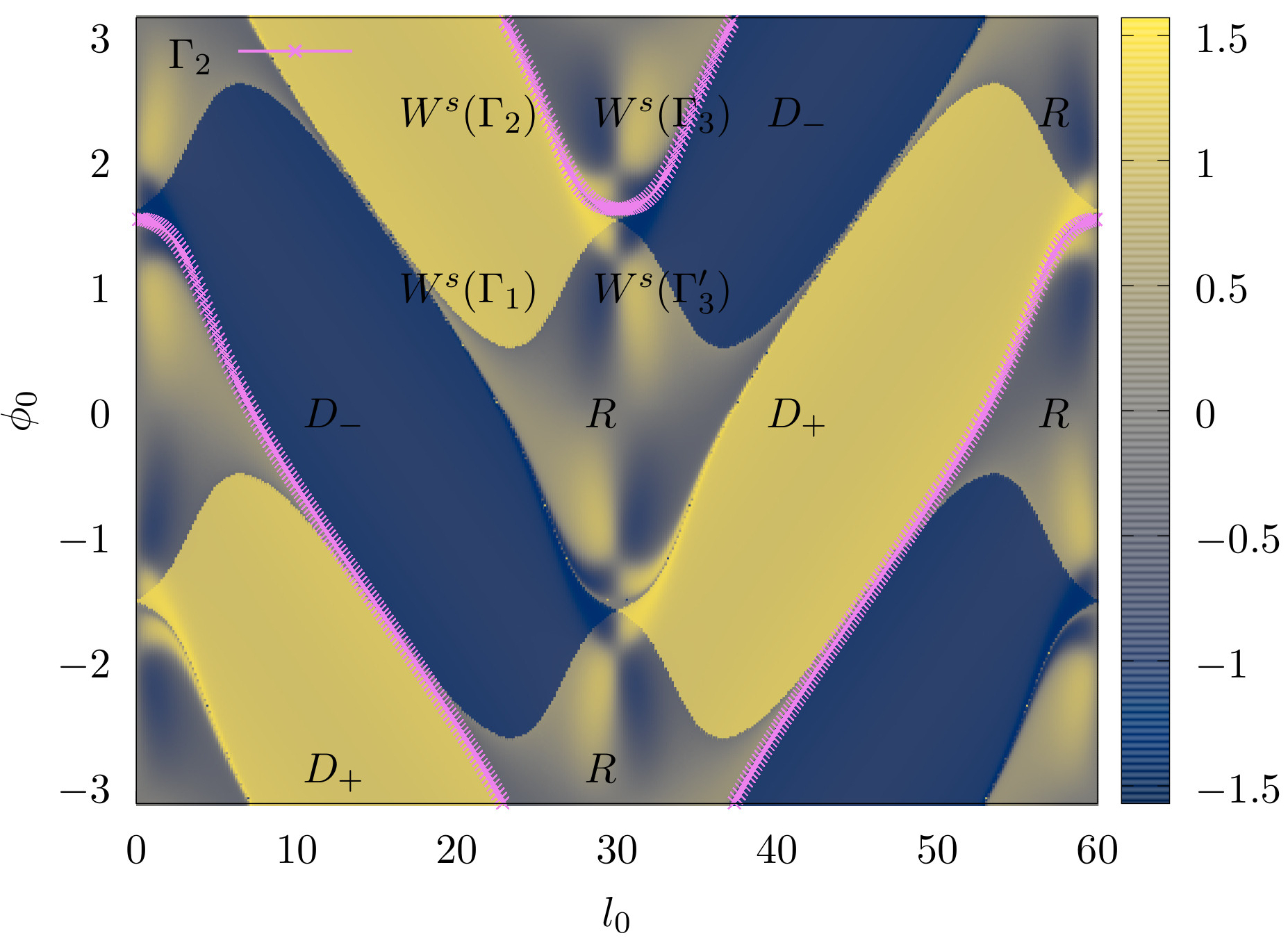}  }}\\
  \subfigure[$ $]{  
 \label{fig:FM_DS_Gamma3}
 \scalebox{0.6}{
 \includegraphics{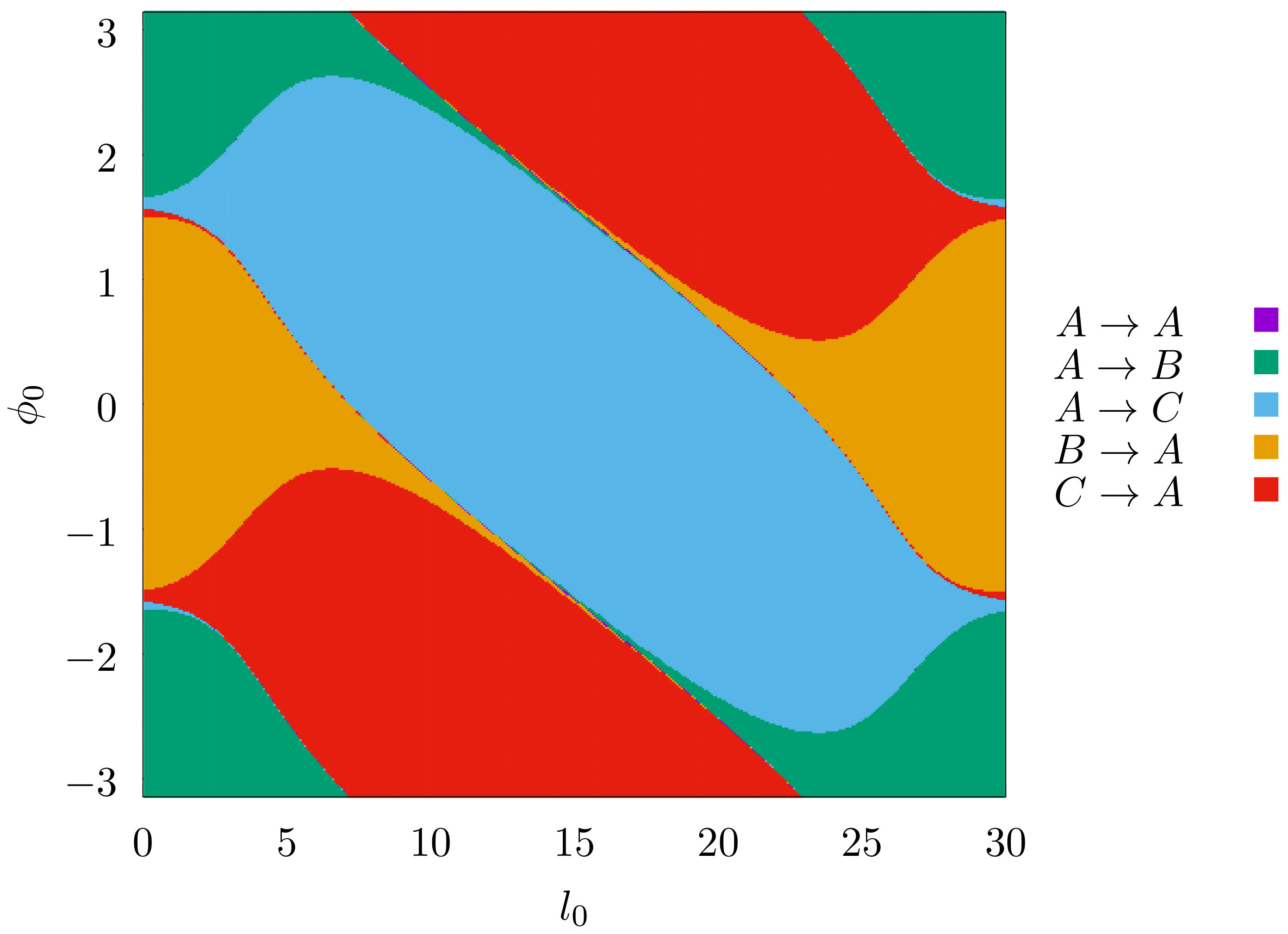}  }}
% &
 \subfigure[$ $ ]{
 \label{fig:LD_DS_Gamma3}
 \scalebox{0.6}{
 \includegraphics{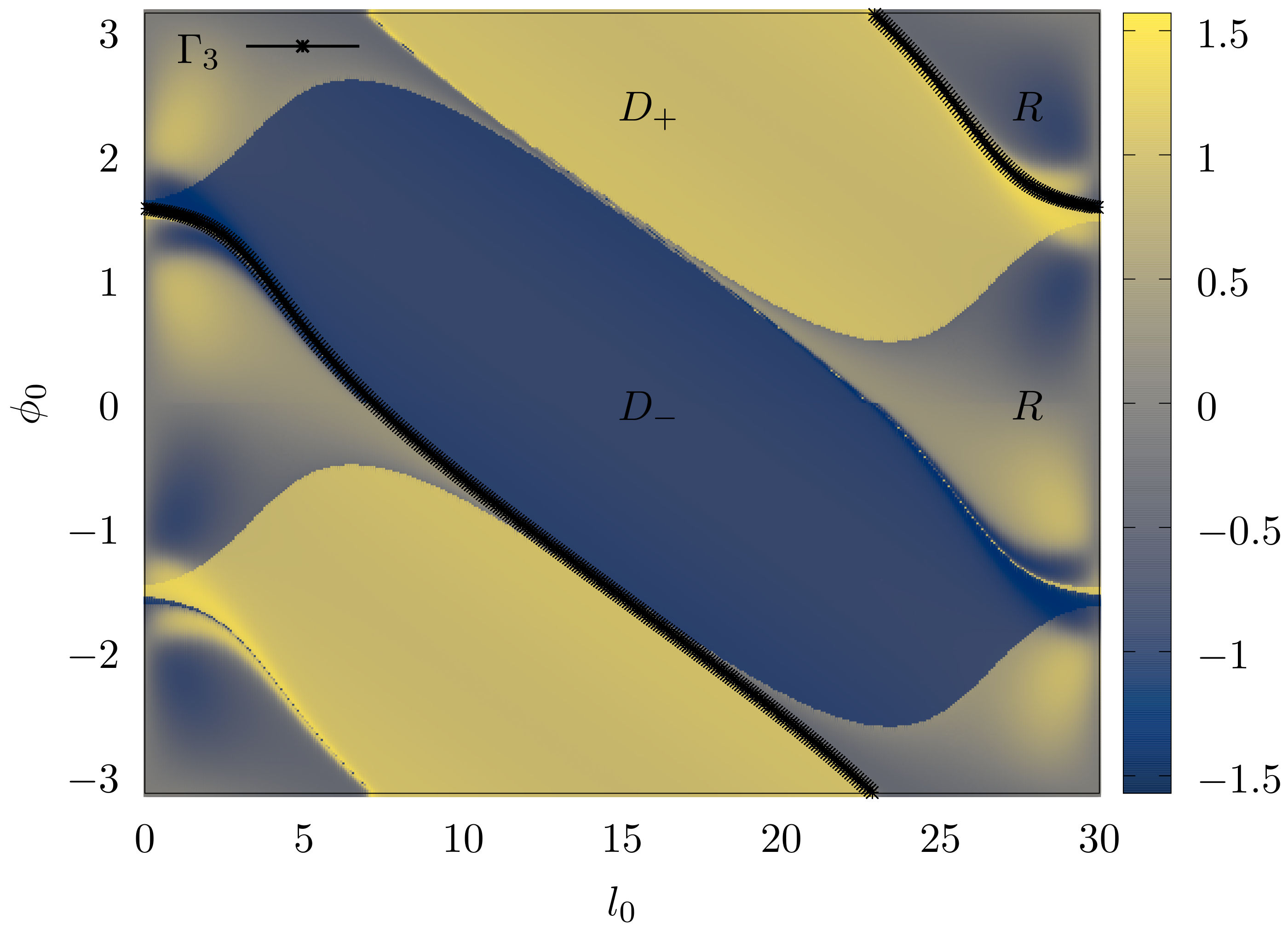}  }}\\

\caption{ Fate map and Lagrangian descriptor evaluated on the dividing surfaces for $\Gamma_1$, $\Gamma_2$, and  $\Gamma_3$.
In order to improve visibility in the Lagrangian descriptor plots, the value $\arctan{(0.005M)}$ is plotted instead of $M$. The intersection between the dividing surface and stable manifolds and unstable manifolds are the curves with a change in the gradient. The yellow and blue curves, where the gradient changes rapidly, are the intersections with the stable and unstable manifolds. There are four large regions in the plot, two regions for dissociation indicated by $D_{-}$ and the other two regions for the inverse process $D_{+}$. The roaming regions are indicated by a letter $R$.
\label{fig:FM_LD}}
\end{center}
\end{figure}

The fate map is an appropriate tool to classify the trajectories associated with rooming and dissociation in phase space. A fate map shows the origin and destiny of the trajectories, considering the evolution for an interval of time. In this case, the origin and destiny of the trajectories are the regions $A$, $B$, and $C$ defined in the previous section. The procedure to calculate the fate map on a dividing surface is the following: consider a point on the dividing surface, integrate forward and backwards the trajectory that crosses this point on the dividing surface until each extreme of the trajectory reaches one of the three regions, and label this point with the corresponding fate. The left side of figure \ref{fig:FM_LD} shows the fate map for each dividing surface. Each different colour in the fate map indicates different transport between regions $A$, $B$, and $C$. The fate maps evaluated on the dividing surfaces of the periodic orbits $\Gamma_1$ and $\Gamma_2$ have all nine possibilities of transport between the regions $A$, $B$, and $C$. In the case for the dividing surface of the periodic orbit $\Gamma_3$, the fate map has only five possibilities.

In the three fate maps there exist red regions ($ C \rightarrow A $), and light blue regions ($ A \rightarrow C$). The light blue regions represent the trajectories that dissociate. The green and orange regions, ($ A \rightarrow B $) and ($ B \rightarrow A $),  are the roaming regions. There are similarities between both of the fate maps for $\Gamma_2$ and $\Gamma_3$. The similarity between the fate maps is a consequence of the proximity between the orbits $\Gamma_3$ and $\Gamma_2$, see figure \ref{fig:orbits_3D}.

In the configuration space, all the trajectories that travel from one well to the other well should cross the projection of the periodic orbits $\Gamma_3$ and $\Gamma_2$. For this reason, the trajectories associated with roaming cross the dividing surface of $\Gamma_3$ and $\Gamma_2$ in the phase space.    

The trajectories that start in one well and escape to the asymptotic region need to cross the projection of $\Gamma_1$ in the configuration space as well. These trajectories are associated with dissociation. The fate map of the dividing surface of $\Gamma_1$ shows which trajectories escape to the asymptotic region and which trajectories enter the interaction region.

%% Invariant manifolds and jumps LD

In order to show the relation between the different trajectories' fates and the stable and unstable manifolds of the unstable periodic orbits $\Gamma_1$, $\Gamma_2$, and $\Gamma_3$ we calculate the Lagrangian descriptors on the dividing surfaces for the three hyperbolic periodic orbits. The same segments of the trajectories are used to calculate the Lagrangian descriptors and the fate maps. Like in the calculation of the fate maps, the values of $\tau_-$ and $\tau_+$ for each trajectory are the times necessary to reach one of the three regions. Once a trajectory reaches one of the three regions, its integration stops. The results for the Lagrangian descriptors are on the right side of figure \ref{fig:FM_LD}. 

The abrupt jumps in the values of the Lagrangian descriptors are related to the different behaviours of the trajectories in a neighbourhood of the stable and unstable manifolds of the unstable periodic orbits in the phase space. The stable and unstable manifolds are surfaces of dimension 2, and their intersections with a dividing surface are curves. There is a complete match between the boundaries of regions in the fate maps and the abrupt changes in the Lagrangian descriptors plots for each dividing surface. This agreement is a manifestation of the division of the phase space into the roaming and dissociation regions by the stable and unstable manifolds.

There are three different types of regions in the plots of the Lagrangian descriptor evaluated on the dividing surface: blue, yellow, and mixed regions. The colours in the Lagrangian descriptor evaluated in the dividing surfaces are related to the fates. The blue regions in the Lagrangian descriptors plots correspond to trajectories that start in one potential well, cross the dividing surface, and finally escape to infinity. These are trajectories associated with dissociation. The yellow regions correspond to trajectories that start in the asymptotic region, cross the dividing surfaces, and finally reach a potential well. The regions with mixed colours in the Lagrangian descriptor plot without abrupt changes are associated with the trajectories that connect the two wells. The mixed colours are related to other segments of the stable and unstable manifolds contained in the same set of initial conditions, and more integration time is necessary to visualise them.

\begin{figure}
 \begin{center}

 \subfigure[$ $ ]{
 \label{fig:FM_DS_Gamma2_zoom}
 \scalebox{0.6}{
 \includegraphics{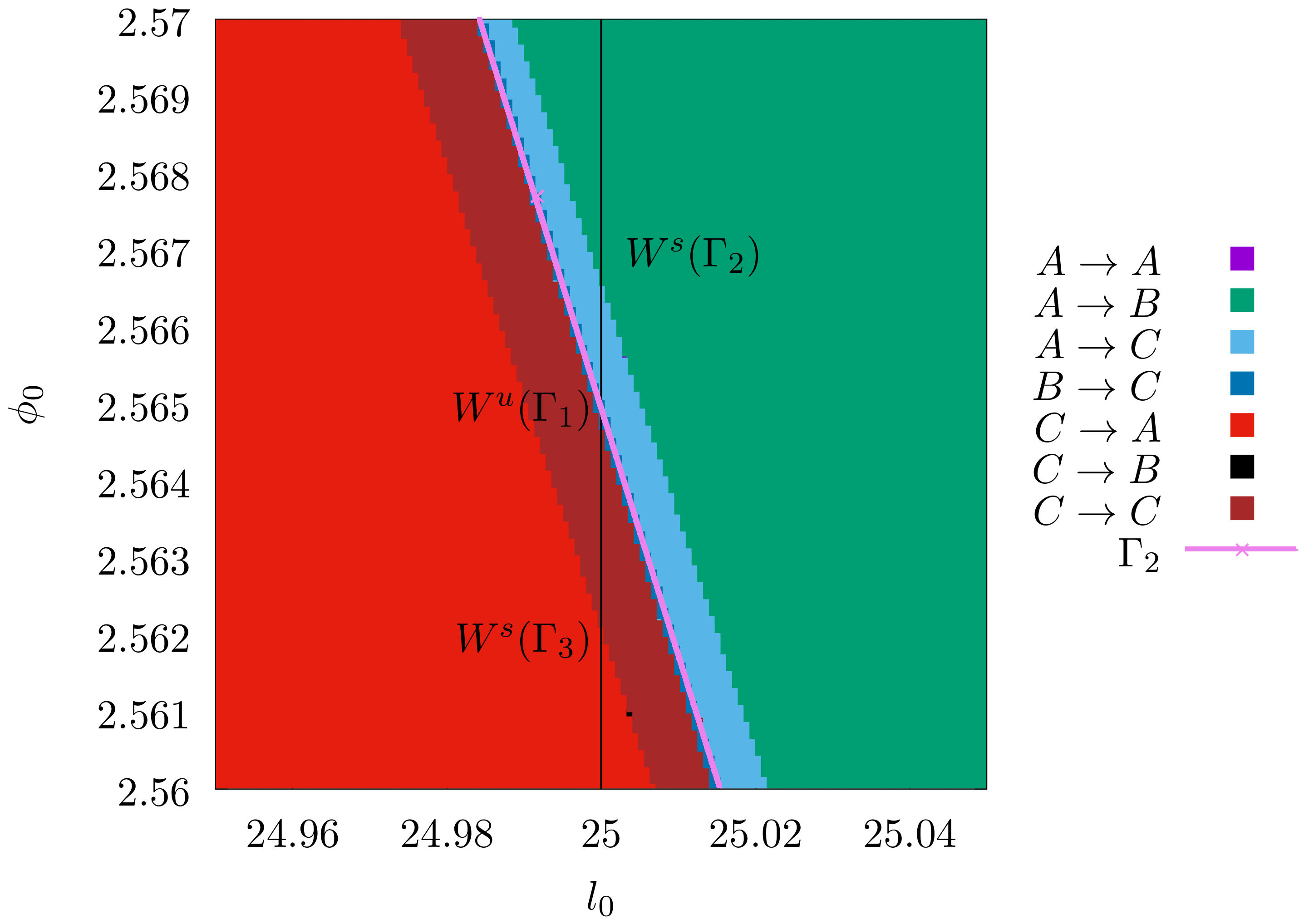}  }}
% &
  \subfigure[$ $]{  
 \label{fig:LD_DS_Gamma_zoom}
 \scalebox{0.6}{
 \includegraphics{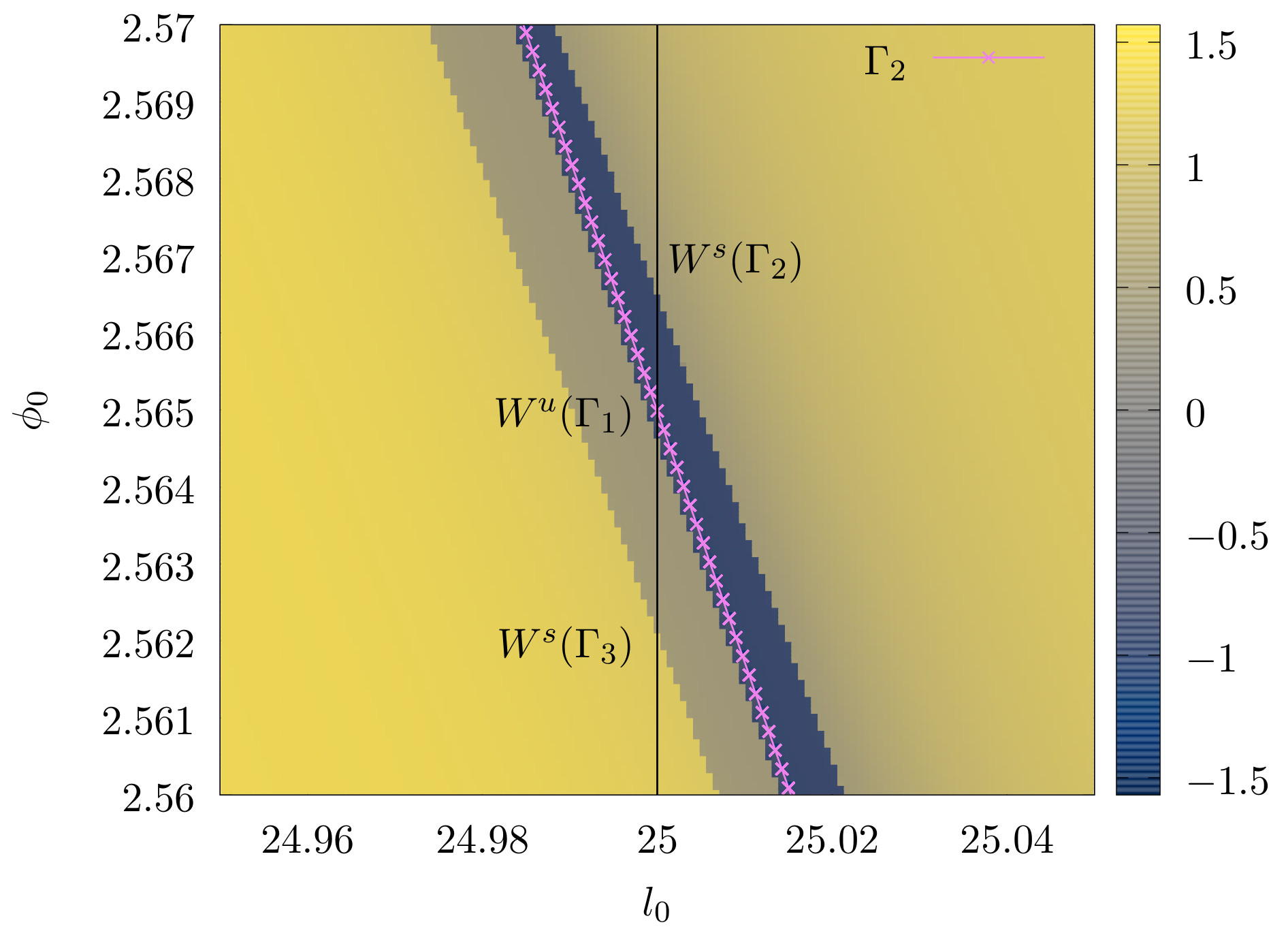}  }}

\caption{ Fate map and Lagrangian descriptor evaluated on dividing surface of orbit $\Gamma_2$ around the point $(l_2=25,\phi_2 = 2.565)$. To improve visibility in the Lagrangian descriptor plot, the value $\arctan{(0.005M)}$ is plotted instead $M$ in \ref{fig:LD_DS_Gamma_zoom}.
\label{fig:FM_LD_zoom}}
\end{center}
\end{figure}

\begin{figure}
 \begin{center}
 
 \subfigure[$W^s(\Gamma_2)$]{  
 \label{fig:orbits_AB_AC}
 \scalebox{0.6}{
 \includegraphics{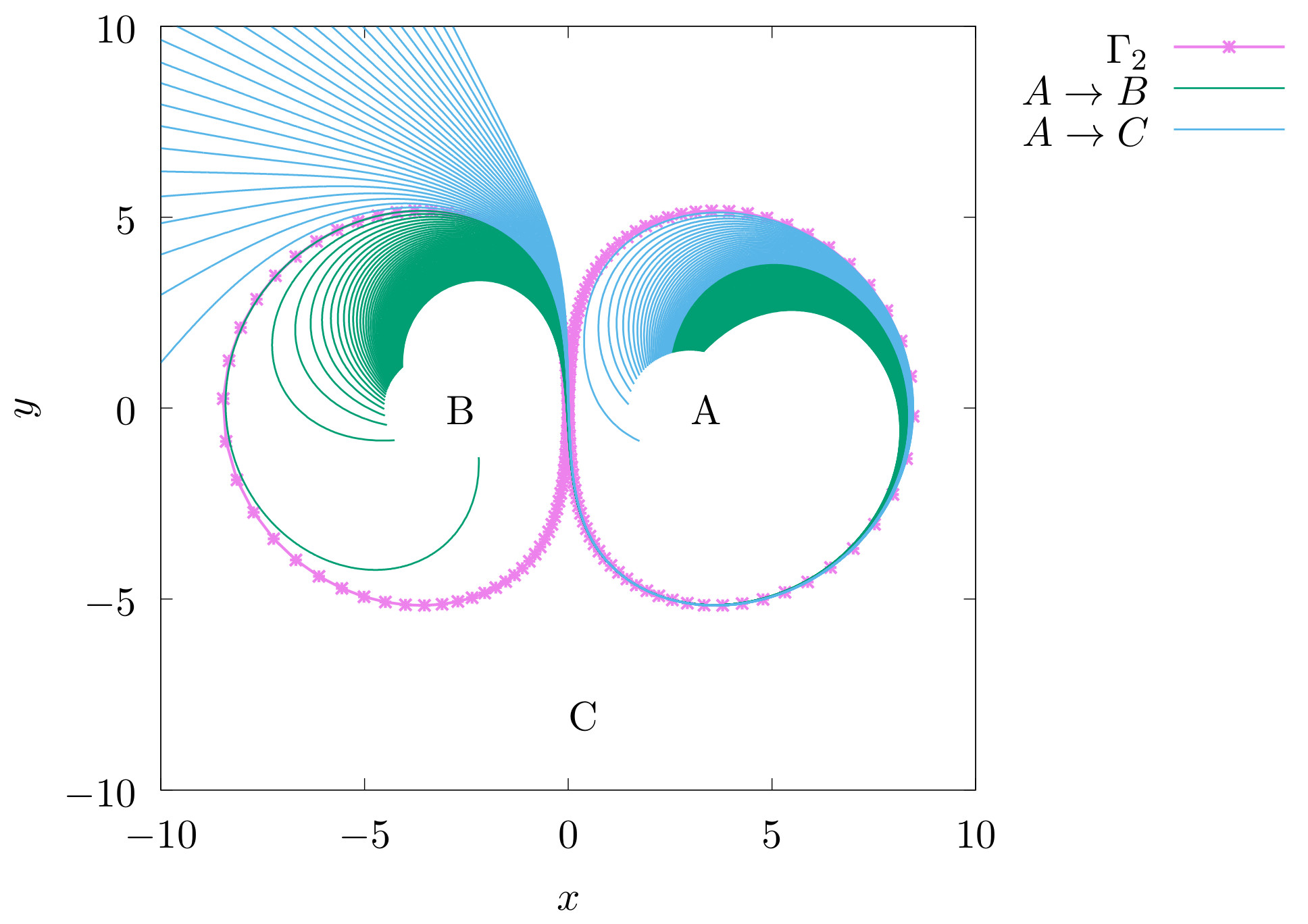}  }}
% &
 \subfigure[ $\Gamma_2$  ]{
 \label{fig:orbits_AC_BC}
 \scalebox{0.6}{
 \includegraphics{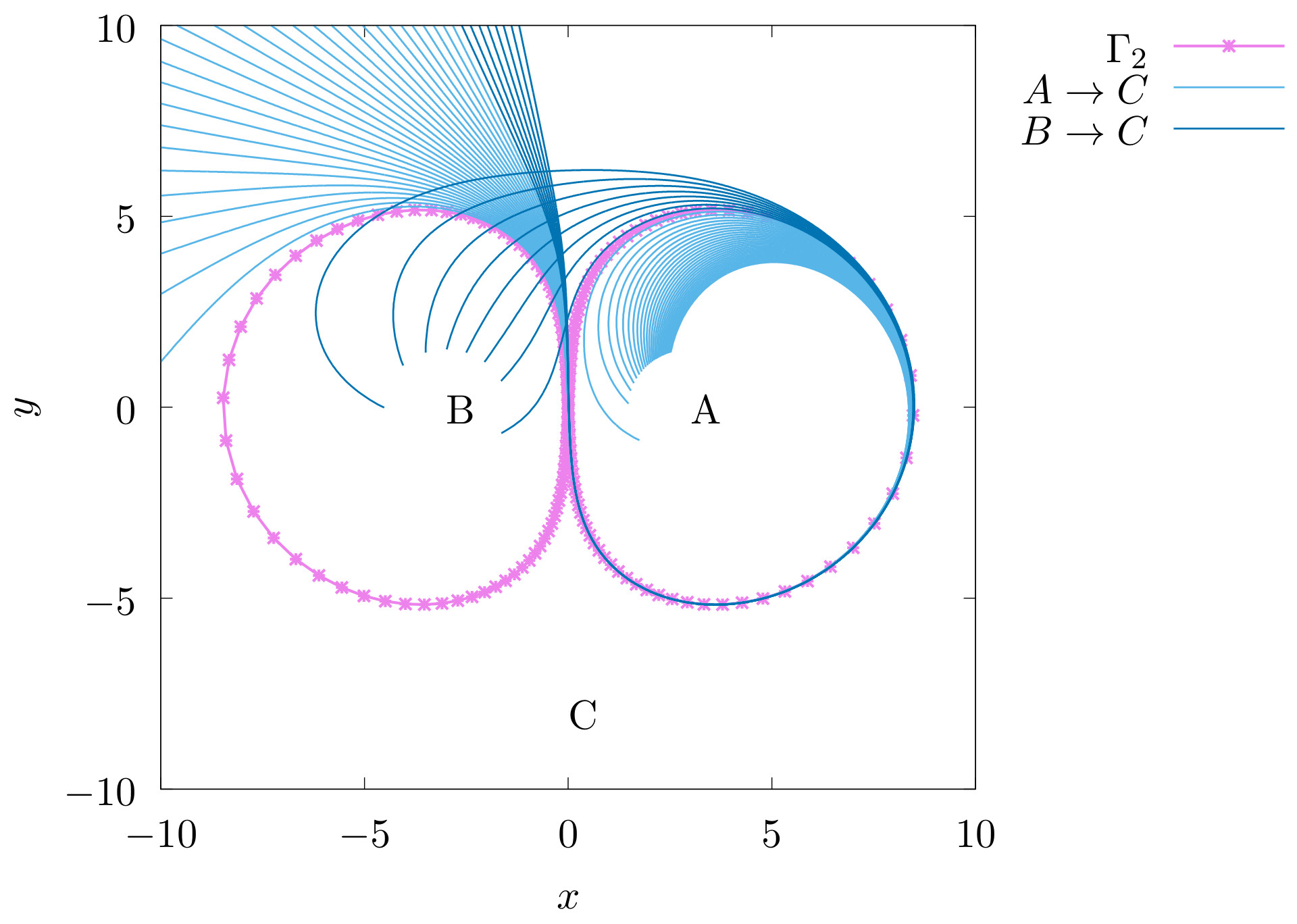}  }}\\
% &
\subfigure[$W^u(\Gamma_1)$]{
 \label{fig:orbits_BC_CC}
 \scalebox{0.6}{
 \includegraphics{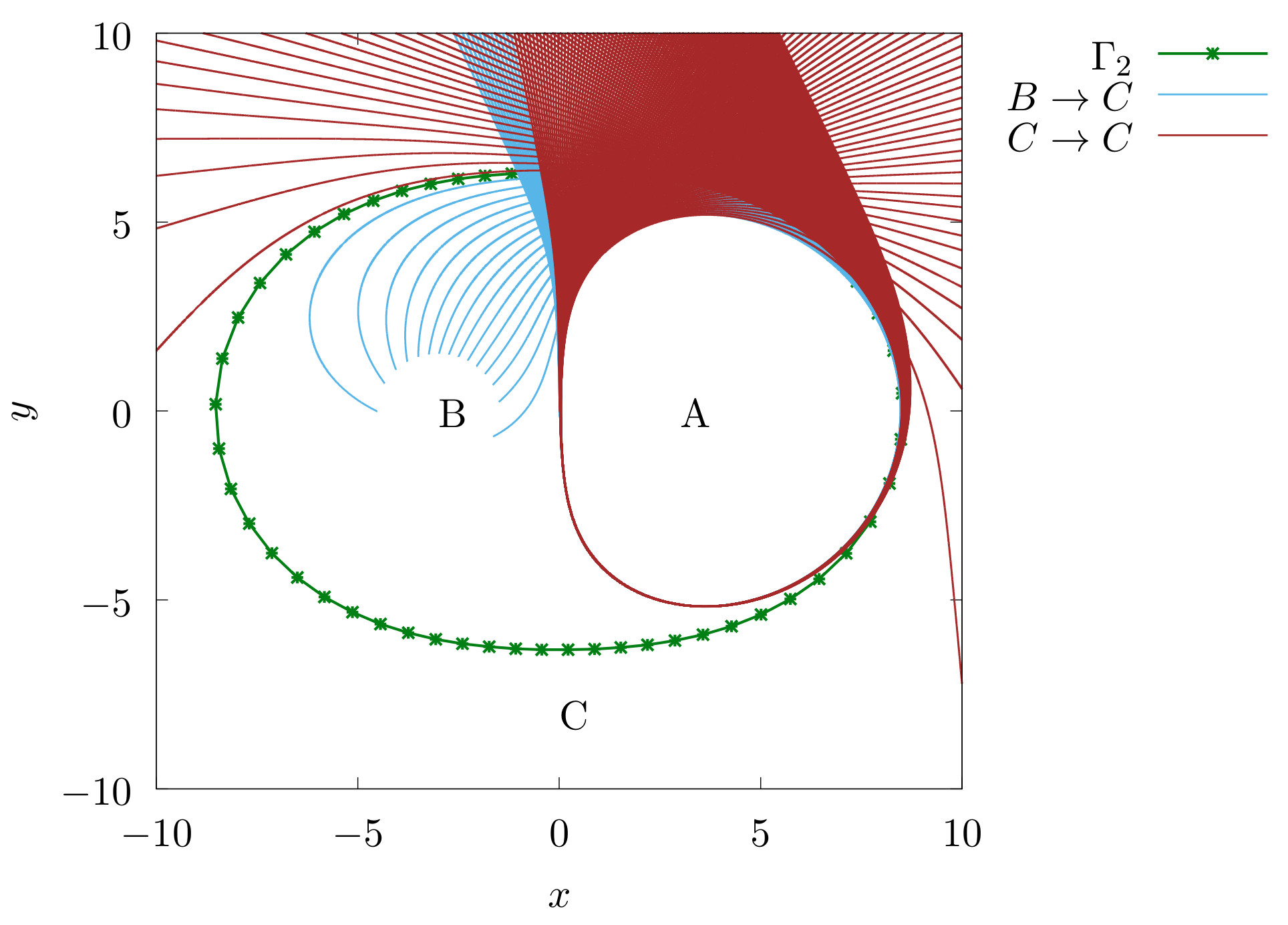}  }}
% & 
\subfigure[$W^s(\Gamma_3)$]{
 \label{fig:orbits_CC_CA}
 \scalebox{0.6}{
 \includegraphics{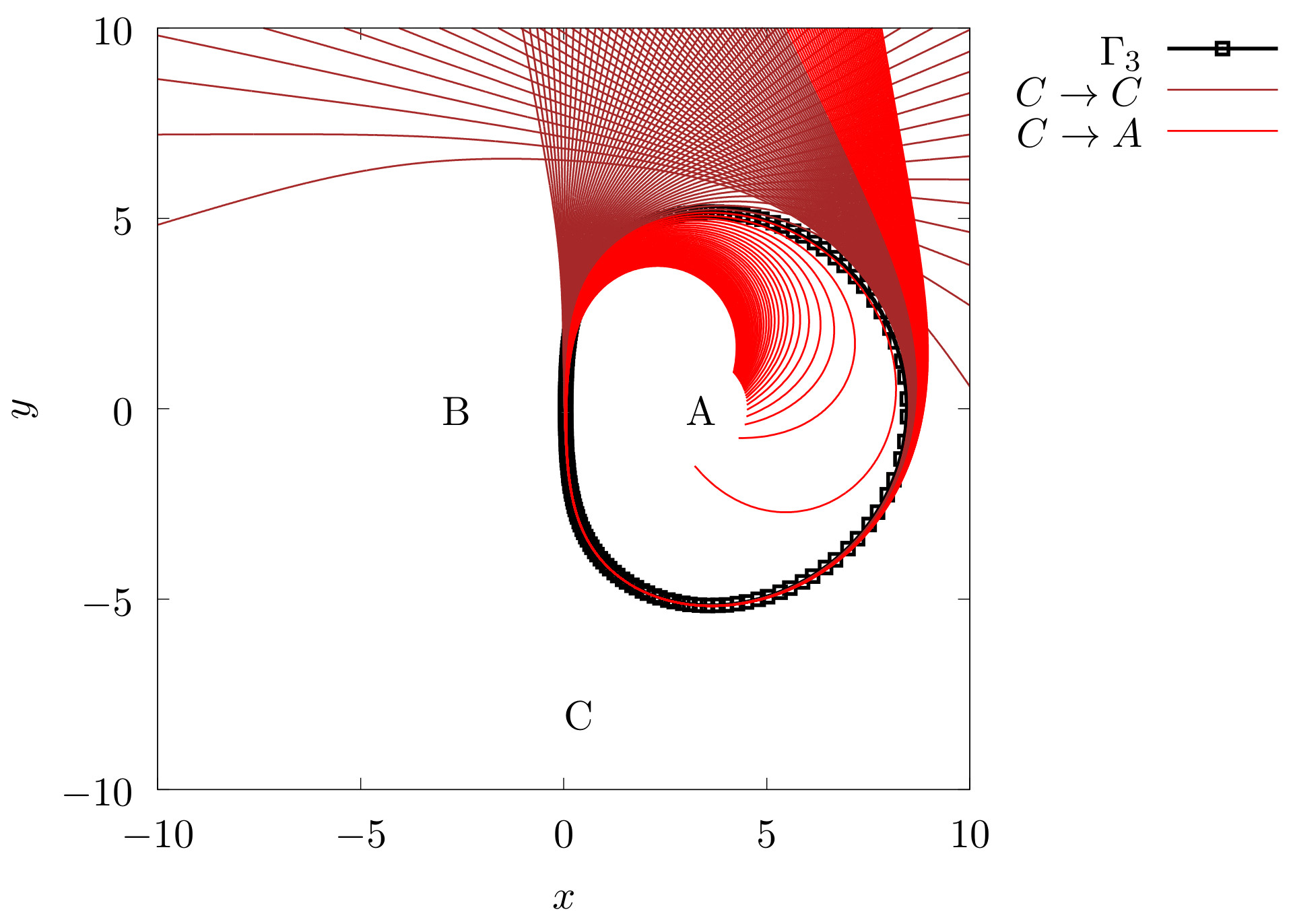} }} \\

\caption{ Projection of the trajectories that cross the black line in figure \ref{fig:FM_DS_Gamma2_zoom} on the dividing surface of $\Gamma_2$. The stable and unstable manifolds are the boundary between two sets of trajectories with a different fate.
\label{fig:orbits}}
\end{center}
\end{figure}

The procedure to identify the stable and unstable manifolds associated with the transport from the Lagrangian descriptor and fate map plots is the following. For illustration, consider magnifications of the plots around the point $(l_2=25,\phi_2 = 2.565)$, close to the periodic orbit $\Gamma_2$, see figure \ref{fig:FM_LD_zoom}. There are fine strips close to $\Gamma_2$ on both magnifications. Then consider the vertical line of initial conditions in figures \ref{fig:LD_DS_Gamma_zoom} and \ref{fig:FM_DS_Gamma2_zoom}, and their trajectories used to calculate the fate map and the Lagrangian descriptor. The projection of these orbits in the configuration space is in figure \ref{fig:orbits}. For example, the trajectory on the boundary between the green region ($A \rightarrow B$) and blue region ($A \rightarrow C$) converges to the invariant manifold $W^s(\Gamma_2)$, see the figures \ref{fig:FM_DS_Gamma2_zoom} and \ref{fig:orbits_AB_AC}. Continuing with the same procedure for the other boundaries is possible to identify the invariant manifolds associated with the boundaries between regions, see  \ref{fig:orbits_AC_BC}, \ref{fig:orbits_BC_CC}, and \ref{fig:orbits_CC_CA}. 

\begin{figure}
 \begin{center}
 
 \subfigure[$ $]{  
 \label{fig:LD_DS_Gamma2_zoom_line}
 \scalebox{0.6}{
 \includegraphics{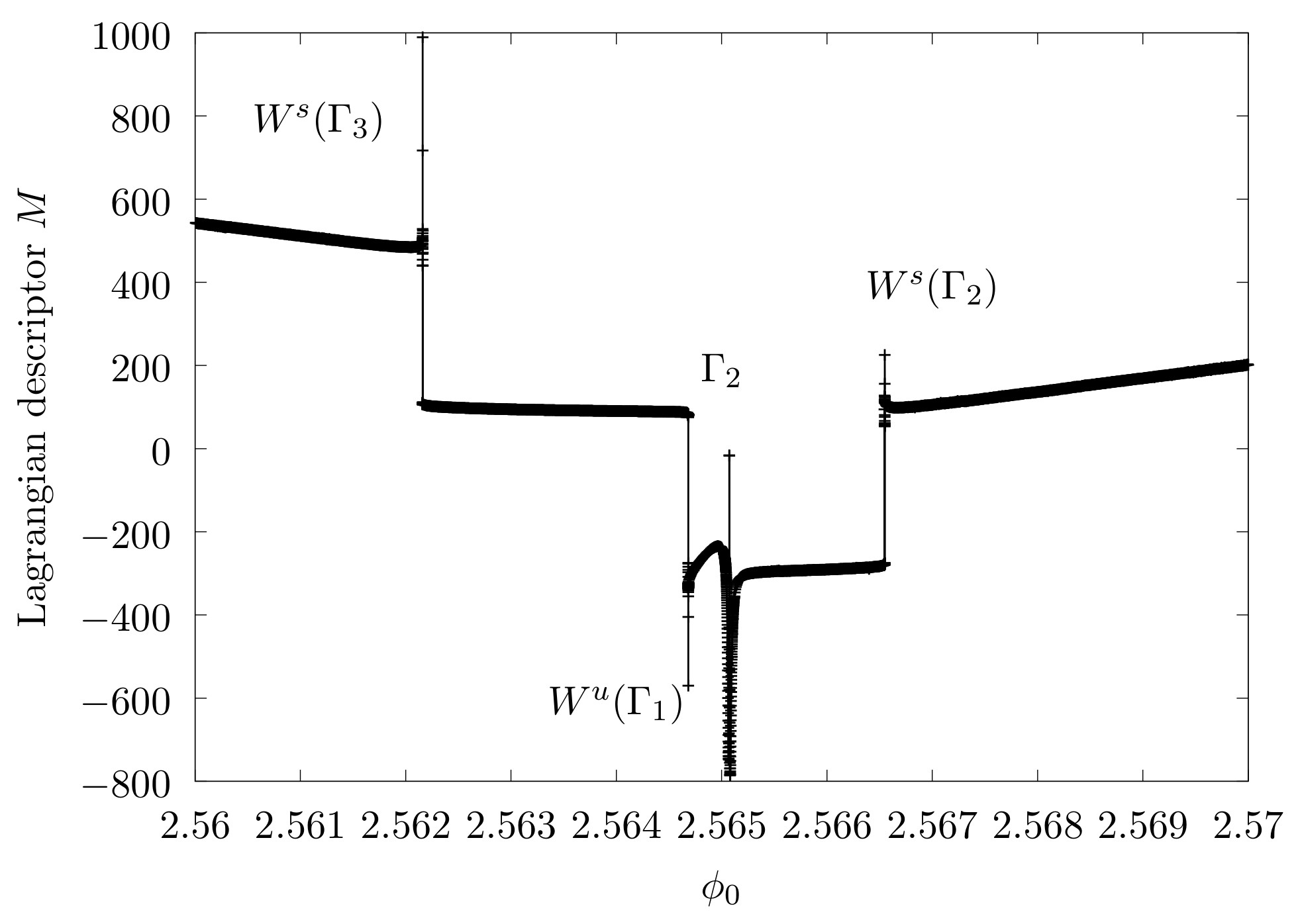}  }}
% &
 \subfigure[$ $ ]{
 \label{fig:LD_DS_Gamma2_zoom_zoom_line}
 \scalebox{0.6}{
 \includegraphics{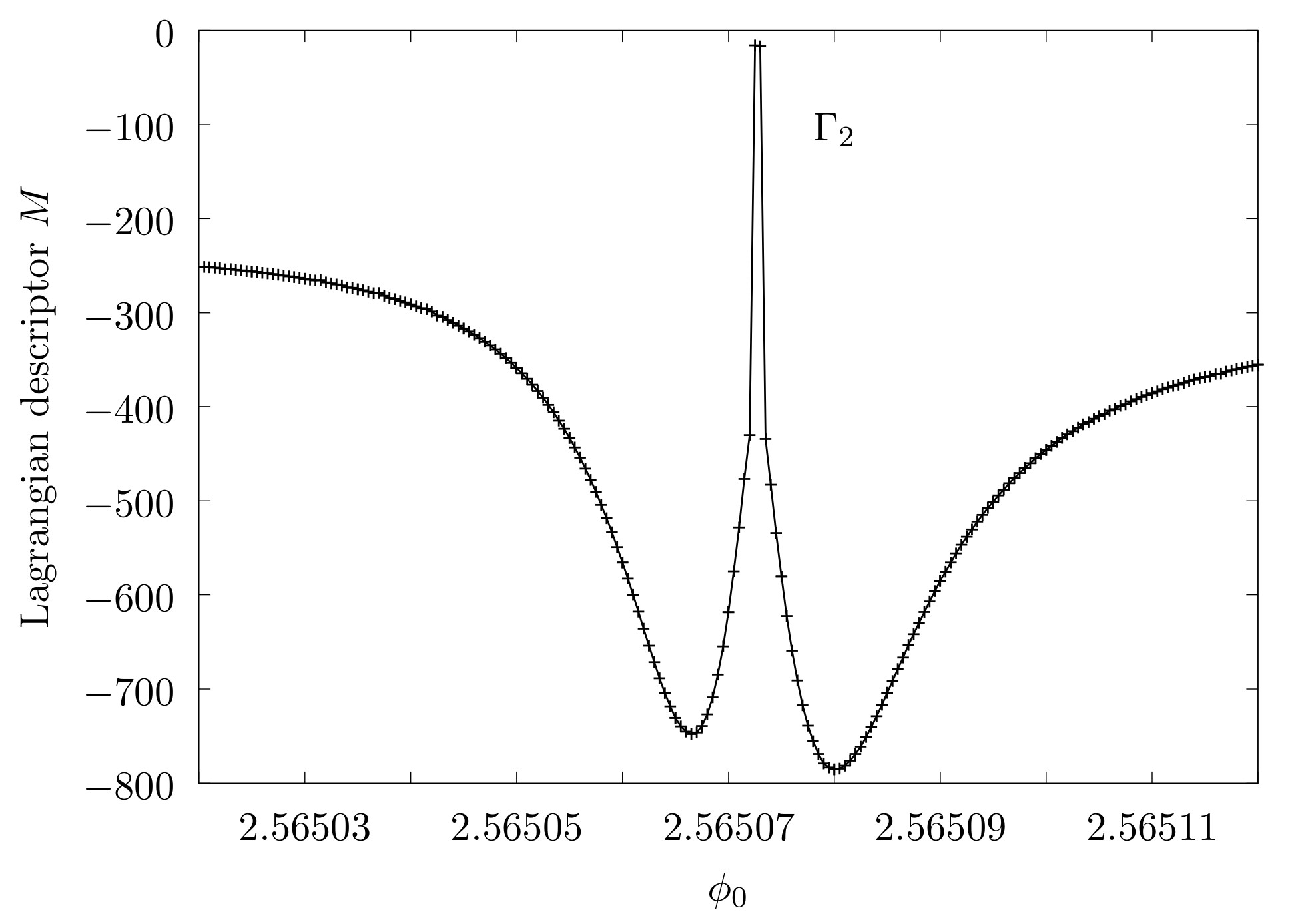}  }}
\\

\caption{ Lagrangian descriptor evaluated on the black vertical line on figure \ref{fig:LD_DS_Gamma_zoom}. Panel \ref{fig:LD_DS_Gamma2_zoom_zoom_line} is a magnification of panel \ref{fig:LD_DS_Gamma2_zoom_line} around the periodic orbit $\Gamma_2$. 
\label{fig:LD_line}}
\end{center}
\end{figure}

\newpage
\section{Conclusions and final remarks}
\label{sec:concl}

The double Morse system is an example of the two centre problem with a rich phase space structure. The periodic orbits type I, II, and III associated with the roaming and the dissociation phenomena are hyperbolic and considerably unstable ($\lambda_{k,Max}>>1$, $k=1,2,3$) for energies close to the asymptotic value of the potential energy.

Lagrangian descriptors are useful tools for finding the invariant manifolds of considerably unstable periodic orbits in a simple way. The abrupt peaks in the Lagrangian descriptor plots reveals the intersections between the set of initial conditions and the stable and unstable manifolds. The set of trajectories that generate the abrupt peaks in the Lagrangian descriptors plots do not necessarily return to the set of initial conditions. This property of the Lagrangian descriptors allows one to easily study the stable and unstable manifolds in the phase space of open systems. The principle that generates those peaks is based on the different behaviour of the trajectories in the stable and unstable manifolds and the trajectories in their neighbourhoods.

In the case of 2 degree of freedom Hamiltonian systems, it is easy to visualise the structure of the tangle between the stable and unstable manifolds using 2 dimensional sets of initial conditions. The stable and unstable manifolds in the constant energy manifolds have 2 dimensions, then their intersection with the set of initial conditions is a set of curves. In Hamiltonian systems with more than 2 degrees of freedom, the sets of initial conditions necessary to completely visualise the structure of the tangles between stable and unstable manifolds of periodic orbits and NHIMs have more dimensions. Their complete visualisation and analysis is a challenging open problem. Some recent works on visualisation and analysis of phase space structures for 3 degree of freedom Hamiltonian systems are  \cite{Gonzalez2012,Drotos2014,Gonzalez2015,Gonzalez2019,GC2011,MK2011,MK20112,GC2013}.

The dividing surfaces of periodic orbits of type I, II, and III are convenient surfaces to study the transport between the potential wells. The fate maps evaluated on the dividing surfaces of those periodic orbits allow one to analyse the transport between the regions that define the dissociation and roaming. The boundaries of the different regions on the fate map correspond to the stable and unstable manifolds of hyperbolic periodic orbits on the Lagrangian descriptor plots. The stable and unstable manifolds of type I, II, and III of periodic orbits determine the dissociation and roaming in this system.

The stable and unstable manifolds of the hyperbolic periodic orbits type I, II, and III are robust under perturbations due to the large instability of the periodic orbits. Therefore, it is possible to observe a similar mechanism for roaming and dissociation in other systems with two wells in the potential energy when the energy is close to the dissociation threshold. 

\section{Acknowledgments}
\label{sec:Acknowledgments}

We acknowledge the support of EPSRC Grant no. EP/P021123/1. S W acknowledges the support of the Office of Naval Research (Grant No.~N00014-01-1-0769).

%\printbibliography

%\bibliography{bibliography}
\bibliography{biblio}

\begin{thebibliography}{10}

\bibitem{carpenter2018dynamics}
Barry~K Carpenter, Gregory~S Ezra, Stavros~C Farantos, Zeb~C Kramer, and
  Stephen Wiggins.
\newblock {Dynamics on the Double Morse Potential: A Paradigm for Roaming
  Reactions with no Saddle Points}.
\newblock {\em Regular and Chaotic Dynamics}, 23(1):60--79, 2018.

\bibitem{townsend2004roaming}
D~Townsend, S~A Lahankar, S~K Lee, S~D Chambreau, A~G Suits, X~Zhang,
  J~Rheinecker, L~B Harding, and J~M Bowman.
\newblock {The roaming atom: Straying from the reaction path in formaldehyde
  decomposition}.
\newblock {\em Science}, 306(5699):1158--1161, 2004.

\bibitem{Bowman2011Suits}
J~M Bowman and A~G Suits.
\newblock {Roaming reaction: The third way}.
\newblock {\em Phys. Today}, 64(11):33, 2011.

\bibitem{Bowman2011roaming}
J~M Bowman and B~C Shepler.
\newblock {Roaming radicals}.
\newblock {\em Ann. Rev. Phys. Chem.}, 62:531--553, 2011.

\bibitem{BowmanRoaming}
J~M Bowman.
\newblock {Roaming}.
\newblock {\em Molecular Physics}, 112(19):2516--2528, 2014.

\bibitem{suits2008}
A~G Suits.
\newblock {Roaming Atoms and Radicals: A New Mechanism in Molecular
  Dissociation}.
\newblock {\em Acc. Chem. Res.}, 41(7):873--881, 2008.

\bibitem{Mauguiere2017}
F~A~L Mauguiere, P~Collins, Z~C Kramer, B~K Carpenter, G~S Ezra, S~C Farantos,
  and S~Wiggins.
\newblock {Roaming: {A} phase space perspective}.
\newblock {\em Ann. Rev. Phys. Chem.}, 2017.

\bibitem{Wiggins_book1994}
S~Wiggins.
\newblock {\em {Normally hyperbolic invariant manifolds in dynamical systems}}.
\newblock Springer-Verlag, New York, 1994.

\bibitem{Wiggins2001}
S~Wiggins, L~Wiesenfeld, C~Jaff{\'{e}}, and T~Uzer.
\newblock {Impenetrable Barriers in Phase-Space}.
\newblock {\em Phys. Rev. Lett.}, 86(24):5478--5481, jun 2001.

\bibitem{Uzer2002}
T~Uzer, Charles Jaff{\'{e}}, Jes{\'{u}}s Palaci{\'{a}}n, Patricia Yanguas, and
  Stephen Wiggins.
\newblock {The geometry of reaction dynamics}.
\newblock {\em Nonlinearity}, 15(4):957, 2002.

\bibitem{Wiggins08}
H~Waalkens, R~Schubert, and S~Wiggins.
\newblock {Wigner's dynamical trasition state theory in phase space: classical
  and quantum}.
\newblock {\em Nonlinearity}, 21:R1--R118, 2008.

\bibitem{Wiggins2016}
Stephen Wiggins.
\newblock {The role of normally hyperbolic invariant manifolds (NHIMS) in the
  context of the phase space setting for chemical reaction dynamics}.
\newblock {\em Regular and Chaotic Dynamics}, 21(6):621--638, 2016.

\bibitem{Mauguiere2015}
Fr{\'{e}}d{\'{e}}ric A~L Maugui{\`{e}}re, Peter Collins, Zeb~C Kramer, Barry~K
  Carpenter, Gregory~S Ezra, Stavros~C Farantos, and Stephen Wiggins.
\newblock {Phase Space Structures Explain Hydrogen Atom Roaming in Formaldehyde
  Decomposition}.
\newblock {\em The Journal of Physical Chemistry Letters}, 6(20):4123--4128,
  2015.

\bibitem{krajvnak2018phase}
Vladim{\'i}r Kraj{\v{n}}{\'a}k and Holger Waalkens.
\newblock {The phase space geometry underlying roaming reaction dynamics}.
\newblock {\em Journal of Mathematical Chemistry}, pages 1--38, 2018.

\bibitem{krajvnak2018influence}
Vladim{\'i}r Kraj{\v{n}}{\'a}k and Stephen Wiggins.
\newblock {Influence of mass and potential energy surface geometry on roaming
  in Chesnavich's CH 4+ model}.
\newblock {\em The Journal of chemical physics}, 149(9):94109, 2018.

\bibitem{chaos}
J~A~J Madrid and A~M Mancho.
\newblock {Distinguished trajectories in time dependent vector fields}.
\newblock {\em Chaos}, 19:13111, 2009.

\bibitem{patra2018detecting}
Sarbani Patra and Srihari Keshavamurthy.
\newblock {Detecting reactive islands using Lagrangian descriptors and the
  relevance to transition path sampling}.
\newblock {\em Physical Chemistry Chemical Physics}, 20(7):4970--4981, 2018.

\bibitem{craven2016deconstructing}
Galen~T Craven and Rigoberto Hernandez.
\newblock {Deconstructing field-induced ketene isomerization through Lagrangian
  descriptors}.
\newblock {\em Physical Chemistry Chemical Physics}, 18(5):4008--4018, 2016.

\bibitem{craven2017lagrangian}
Galen~T Craven, Andrej Junginger, and Rigoberto Hernandez.
\newblock {Lagrangian descriptors of driven chemical reaction manifolds}.
\newblock {\em Physical Review E}, 96(2):22222, 2017.

\bibitem{craven2015lagrangian}
Galen~T Craven and Rigoberto Hernandez.
\newblock {Lagrangian descriptors of thermalized transition states on
  time-varying energy surfaces}.
\newblock {\em Physical review letters}, 115(14):148301, 2015.

\bibitem{junginger2016transition}
Andrej Junginger, Galen~T Craven, Thomas Bartsch, F~Revuelta, F~Borondo, R~M
  Benito, and Rigoberto Hernandez.
\newblock {Transition state geometry of driven chemical reactions on
  time-dependent double-well potentials}.
\newblock {\em Physical Chemistry Chemical Physics}, 18(44):30270--30281, 2016.

\bibitem{revuelta2017transition}
F~Revuelta, Galen~T Craven, Thomas Bartsch, F~Borondo, R~M Benito, and
  Rigoberto Hernandez.
\newblock {Transition state theory for activated systems with driven anharmonic
  barriers}.
\newblock {\em The Journal of chemical physics}, 147(7):74104, 2017.

\bibitem{junginger2017chemical}
Andrej Junginger, Lennart Duvenbeck, Matthias Feldmaier, J{\"{o}}rg Main,
  G{\"{u}}nter Wunner, and Rigoberto Hernandez.
\newblock {Chemical dynamics between wells across a time-dependent barrier:
  Self-similarity in the Lagrangian descriptor and reactive basins}.
\newblock {\em The Journal of chemical physics}, 147(6):64101, 2017.

\bibitem{feldmaier2017obtaining}
Matthias Feldmaier, Andrej Junginger, J{\"{o}}rg Main, G{\"{u}}nter Wunner, and
  Rigoberto Hernandez.
\newblock {Obtaining time-dependent multi-dimensional dividing surfaces using
  Lagrangian descriptors}.
\newblock {\em Chemical Physics Letters}, 687:194--199, 2017.

\bibitem{JH2015}
A~Junginger and R~Hernandez.
\newblock {Uncovering the Geometry of Barrierless Reactions Using {L}agrangian
  Descriptors}.
\newblock {\em J. Phys. Chem. B}, 2015.

\bibitem{lopesino2017}
Carlos Lopesino, Francisco Balibrea-Iniesta, V{\'{i}}ctor {Garc{\'{i}}a
  Garrido}, Stephen Wiggins, and A~Mancho.
\newblock {A Theoretical Framework for Lagrangian Descriptors}.
\newblock {\em International Journal of Bifurcation and Chaos}, 27:1730001,
  2017.

\bibitem{balibrea2016lagrangian}
Francisco Balibrea-Iniesta, Carlos Lopesino, Stephen Wiggins, and Ana~M Mancho.
\newblock {Lagrangian Descriptors for Stochastic Differential Equations: A Tool
  for Revealing the Phase Portrait of Stochastic Dynamical Systems}.
\newblock {\em International Journal of Bifurcation and Chaos}, 26(13):1630036,
  2016.

\bibitem{mmw14}
C~Mendoza, A~M Mancho, and S~Wiggins.
\newblock {Lagrangian Descriptors and the Assesment of the Predictive Capacity
  of Oceanic Data Sets.}
\newblock {\em Nonlin. Proc. Geophys.}, 21:677--689, 2014.

\bibitem{Mancho2013}
Ana~M Mancho, Stephen Wiggins, Jezabel Curbelo, and Carolina Mendoza.
\newblock {Lagrangian descriptors: A method for revealing phase space
  structures of general time dependent dynamical systems}.
\newblock {\em Communications in Nonlinear Science and Numerical Simulation},
  18(12):3530--3557, 2013.

\bibitem{Morse29}
P~M Morse.
\newblock {Diatomic Molecules According to the Wave Mechanics. II. Vibrational
  Levels}.
\newblock {\em Physical Review}, 34:57--64, jul 1929.

\bibitem{Kovacks2001}
Z~Kov{\'{a}}cs and L~Wiesenfeld.
\newblock {Topological aspects of chaotic scattering in higher dimensions}.
\newblock {\em Phys. Rev. E}, 63(5):56207, apr 2001.

\bibitem{Tel_book}
Ying-Cheng Lai and Tam{\'{a}}s T{\'{e}}l.
\newblock {\em {Transient Chaos}}.
\newblock Springer-Verlag New York, 2011.

\bibitem{Tel2015}
T{\'{a}}mas T{\'{e}}l.
\newblock {The joy of transient chaos}.
\newblock {\em Chaos: An Interdisciplinary Journal of Nonlinear Science},
  25(9):97619, 2015.

\bibitem{Sanjuan2013}
Jes{\'{u}}s~M Seoane and Miguel A~F Sanju{\'{a}}n.
\newblock {New developments in classical chaotic scattering}.
\newblock {\em Reports on Progress in Physics}, 76(1):16001, 2013.

\bibitem{Zapfe2010}
C~Jung, O~Merlo, T~H Seligman, and W~P~K Zapfe.
\newblock {The chaotic set and the cross section for chaotic scattering beyond
  two degrees of freedom}.
\newblock {\em New Journal of Physics}, 12, 2010.

\bibitem{Drotos2014}
G{\'{a}}bor Dr{\'{o}}tos, Francisco {Gonz{\'{a}}lez Montoya}, Christof Jung,
  and Tam{\'{a}}s T{\'{e}}l.
\newblock {Asymptotic observability of low-dimensional powder chaos in a
  three-degrees-of-freedom scattering system}.
\newblock {\em Phys. Rev. E}, 90(2):22906, aug 2014.

\bibitem{Gonzalez2012}
F~Gonzalez and C~Jung.
\newblock {Rainbow singularities in the doubly differential cross section for
  scattering off a perturbed magnetic dipole}.
\newblock {\em Journal of Physics A: Mathematical and Theoretical},
  45(26):265102, 2012.

\bibitem{Newton_book}
Roger~G Newton.
\newblock {\em {Scattering Theory of Waves and Particles}}.
\newblock McGraw-Hill, 1966.

\bibitem{Perez2019}
Jorge~A P{\'{e}}rez-Hern{\'{a}}ndez and Luis Benet.
\newblock {PerezHz/TaylorIntegration.jl: TaylorIntegration v0.4.1}.
\newblock \url{https://doi.org/10.5281/zenodo.2562353}, feb 2019.

\bibitem{Lega2016}
E~Lega, M~Guzzo, and C~Froeschl{\'{e}}.
\newblock {Theory and Applications of the Fast Lyapunov Indicator (FLI) Method.
  In: Skokos C., Gottwald G., Laskar J. (eds) Chaos Detection and
  Predictability}.
\newblock {\em Lecture Notes in Physics}, (915), 2016.

\bibitem{Cicotta2016}
P~M Cincotta and C~M Giordano.
\newblock {Theory and Applications of the Mean Exponential Growth Factor of
  Nearby Orbits (MEGNO) Method. In: Skokos C., Gottwald G., Laskar J. (eds)
  Chaos Detection and Predictability}.
\newblock {\em Lecture Notes in Physics}, (915), 2016.

\bibitem{Skokos2016}
C~Skokos and {Manos T}.
\newblock {The Smaller (SALI) and the Generalized (GALI) Alignment Indices:
  Efficient Methods of Chaos Detection. In: Skokos C., Gottwald G., Laskar J.
  (eds) Chaos Detection and Predictability}.
\newblock {\em Lecture Notes in Physics}, (915), 2016.

\bibitem{Rackauckas2017}
C~Rackauckas and Q~Nie.
\newblock {DifferentialEquations.jl - A Performant and Feature-Rich Ecosystem
  for Solving Differential Equations in Julia}.
\newblock {\em Journal of Open Research Software}, (5):15, 2017.

\bibitem{Pechukas1979}
Philip Pechukas and Eli Pollak.
\newblock {Classical transition state theory is exact if the transition state
  is unique}.
\newblock {\em The Journal of Chemical Physics}, 71(5):2062--2068, 1979.

\bibitem{Waalkens_2004}
H~Waalkens and S~Wiggins.
\newblock {Direct construction of a dividing surface of minimal flux for
  multi-degree-of-freedom systems that cannot be recrossed}.
\newblock {\em Journal of Physics A: Mathematical and General},
  37(35):L435--L445, aug 2004.

\bibitem{Rafa2019}
R~Garcia-Meseguer, B~K Carpenter, and S~Wiggins.
\newblock {The influence of the solvent's mass on the location of the dividing
  surface for a model Hamiltonian}.
\newblock {\em Chemical Physics Letters: X}, 3:100030, 2019.

\bibitem{Collins_2015}
Fr{\'{e}}d{\'{e}}ric Maugui{\`{e}}re, Peter Collins, Zeb Kramer, Barry
  Carpenter, Gregory Ezra, Stavros Farantos, and Stephen Wiggins.
\newblock {Phase space barriers and dividing surfaces in the absence of
  critical points of the potential energy: Application to roaming in ozone}.
\newblock {\em The Journal of Chemical Physics}, 144, 2015.

\bibitem{Gonzalez2015}
F~Gonzalez and C~Jung.
\newblock {Visualizing the perturbation of partial integrability}.
\newblock {\em Journal of Physics A: Mathematical and Theoretical},
  48(43):435101, oct 2015.

\bibitem{Gonzalez2019}
Francisco Gonzalez, Florentino Borondo, and Christof Jung.
\newblock {Atom scattering off a vibrating surface: An example of chaotic
  scattering with three degrees of freedom}.
\newblock arXiv 1912.05597, 2019.

\bibitem{GC2011}
M~Katsanikas, P~A Patsis, and G~Contopoulos.
\newblock {The structure and evolution of confined tori near a Hamiltonian Hopf
  bifurcation}.
\newblock {\em International Journal of Bifurcation and Chaos},
  21(08):2321--2330, aug 2011.

\bibitem{MK2011}
M~Katsanikas, P~A Patsis, and A~D Pinotsis.
\newblock {Chains of rotational tori and filamentary structures close to high
  multiplicity periodic orbits in a 3d galactic potencial}.
\newblock {\em International Journal of Bifurcation and Chaos},
  21(08):2331--2342, aug 2011.

\bibitem{MK20112}
M~Katsanikas and P~A Patsis.
\newblock {The structure of invariant tori in a 3d galactic potencial}.
\newblock {\em International Journal of Bifurcation and Chaos},
  21(02):467--496, 2011.

\bibitem{GC2013}
M~Katsanikas, P~A Patsis, and G~Contopoulos.
\newblock {Instabilities and stickness in a 3d rotating galactic potencial}.
\newblock {\em International Journal of Bifurcation and Chaos}, 23(02):1330005,
  feb 2013.

\end{thebibliography}

\end{document}